\documentclass[12pt]{iopart}
\usepackage{graphicx}
\expandafter\let\csname equation*\endcsname\relax
\expandafter\let\csname endequation*\endcsname\relax
\usepackage{amsmath}

\usepackage{color,soul}

\usepackage{algorithm}
\usepackage{algpseudocode}
\usepackage{bm}
\usepackage{esvect}
\usepackage{accents}
\usepackage{cite}
\newcommand*{\dt}[1]{%
  \accentset{\mbox{\large\bfseries .}}{#1}}

\begin{document}

\title{Result of the MICROSCOPE Weak Equivalence Principle test}

\author{Pierre Touboul$^1$\footnote{Deceased in February 2021}, Gilles M\'etris$^2$, Manuel Rodrigues$^3$, Joel Berg\'e$^3$, Alain Robert$^4$, Quentin Baghi$^{2,3}$\footnote{Current address: CEA, Centre de Saclay, IRFU/DPhP, 91191 Gif-sur-Yvette, France}, Yves Andr\'e$^4$, Judicael Bedouet$^5$, Damien Boulanger$^3$, Stefanie Bremer$^6$\footnote{Current address: DLR, Institute for Satellite Geodesy and Inertial Sensing, Am Fallturm 9, D-28359 Bremen, Germany}, Patrice Carle$^1$, Ratana Chhun$^3$, Bruno Christophe$^3$, Valerio Cipolla$^4$, Thibault Damour$^7$, Pascale Danto$^4$, Louis Demange$^2$, Hansjoerg Dittus$^8$, Oc\'eane Dhuicque$^3$, Pierre Fayet$^9$, Bernard Foulon$^3$, Pierre-Yves Guidotti$^4$\footnote{Current address: AIRBUS Defence and Space, F-31402 Toulouse, France}, Daniel Hagedorn$^{10}$, Emilie Hardy$^3$, Phuong-Anh Huynh$^3$, Patrick Kayser$^3$, Stephanie Lala$^1$, Claus L\"ammerzahl$^6$, Vincent Lebat$^3$, Fran\c{c}oise Liorzou$^3$, Meike List$^6$\footnote{Current address: DLR, Institute for Satellite Geodesy and Inertial Sensing, Am Fallturm 9, D-28359 Bremen, Germany}, Frank L\"offler$^{10}$, Isabelle Panet$^{11}$, Martin Pernot-Borr\`as$^3$, Laurent Perraud$^4$, Sandrine Pires$^{12}$, Benjamin Pouilloux$^4$\footnote{Current address: KINEIS, F-31520 Ramonville Saint-Agne, France}, Pascal Prieur$^4$, Alexandre Rebray$^3$, Serge Reynaud$^{13}$, Benny Rievers$^6$, Hanns Selig$^6$\footnote{Current address: GERADTS GMBH, Kleiner Ort 8, D-28357 Bremen, Germany}, Laura Serron$^2$, Timothy Sumner$^{14}$, Nicolas Tanguy$^3$, Patrizia Torresi$^4$, Pieter Visser$^{15}$}

\address{$^1$ ONERA, Universit\'e Paris Saclay, F-91123 Palaiseau, France}
\address{$^2$ Universit\'e C\^ote d'Azur, Observatoire de la C\^ote d'Azur, CNRS, IRD, G\'eoazur, 250 avenue Albert Einstein, F-06560 Valbonne, France}
\address{$^3$ ONERA, Universit\'e Paris Saclay, F-92322 Ch\^atillon, France}
\address{$^4$ CNES Toulouse, 18 avenue Edouard Belin - 31401 Toulouse Cedex 9, France}
\address{$^5$ ONERA, Universit\'e de Toulouse, F-31055 Toulouse, France}
\address{$^6$ ZARM, Center of Applied Space Technology and Microgravity, University of Bremen, Am Fallturm, D-28359 Bremen, Germany}
\address{$^7$ IHES, Institut des Hautes Etudes Scientifiques, 35 Route de Chartres, 91440 Bures-sur-Yvette, France}
\address{$^8$ DLR, K\"oln headquarters, Linder H\"ohe, 51147 K\"oln, Germany}
\address{$^9$ Laboratoire de physique de l'Ecole normale sup\'erieure, ENS, Universit\'e PSL, CNRS, Sorbonne Universit\'e, Universit\'e Paris Cit\'e, F-75005 Paris, France, and CPhT, Ecole polytechnique, IPP, Palaiseau, France}
\address{$^{10}$ PTB, Physikalisch-Technische Bundesanstalt, Bundesallee 100, 38116 Braunschweig, Germany}
\address{$^{11}$ IPGP, 35 rue H\'el\`ene Brion, 75013 Paris, France}
\address{$^{12}$ Universit\'e Paris-Saclay, Universit\'e Paris Cit\'e, CEA, CNRS, AIM Paris-Saclay, F-91190 Gif-sur-Yvette, France}
\address{$^{13}$ Laboratoire Kastler Brossel, Sorbonne Universit\'e, CNRS, ENS-PSL Universit\'e, Coll\`ege de France, 75252 Paris, France}
\address{$^{14}$ Blackett Laboratory, Imperial College London, Prince Consort Road, London. SW7 2AZ, United Kingdom}
\address{$^{15}$ Faculty of Aerospace Engineering, Delft University of Technology, Kluyverweg 1, 2629 HS Delft, Netherlands} 

\ead{gilles.metris@oca.eu, manuel.rodrigues@onera.fr}
\vspace{10pt}
\begin{indented}
\item[]September 2022
\end{indented}

\begin{abstract}
  The space mission MICROSCOPE dedicated to the test of the Equivalence Principle (EP) operated from April 25, 2016 {until} the deactivation of the satellite on October 16, 2018. {In this analysis we compare the free-fall accelerations ($a_{\rm A}$ and $a_{\rm B}$) of two test masses in terms of  the E\"otv\"os parameter $\eta({\rm{A, B}}) =  2 \frac{a_{\rm A}- a_{\rm B}}{a_{\rm A}+ a_{\rm B}}$.}  {No EP violation} has been detected for two test masses, {made from platinum  and titanium alloys, in a sequence} of 19 segments lasting from 13 to 198 hours {down to} the limit of the statistical error which is smaller than $10^{-14}$ for $ \eta({\rm{Ti, Pt}})$. {Accumulating} data from all segments leads to $\eta({\rm{Ti, Pt}}) =[-1.5\pm{}2.3{\rm (stat)}\pm{}1.5{\rm (syst)}] \times{}10^{-15}$ {showing no EP}   violation at the level of $2.7\times{}10^{-15}$ {if we combine stochastic and systematic errors quadratically}. This represents an improvement of almost two orders of magnitude with respect to {the previous best such test performed by the E\"ot-Wash group.} {The reliability of this limit}  has been verified by comparing the free falls of two test masses of the same composition (platinum) leading to a null E\"otv\"os parameter with {a statistical uncertainty} of $1.1\times{}10^{-15}$.
\end{abstract}

%

\submitto{\CQG}
%
%
%

\section{Introduction}
{

{The Equivalence Principle (EP) is the foundation stone on which Einstein built his new
theory of gravitation, General Relativity (GR) \cite{einstein08, einstein16}. GR has become 
an essential element in our description of the macroscopic universe, from the big bang to 
black holes and gravitational waves. GR has passed with flying colours many stringent experimental
tests (for reviews, see, e.g., \cite{will14} and chapter 21 in \cite{ParticleDataGroup:2020ssz}).

However, fundamental physics is facing several conundrums which suggest the need to extend
our present theoretical framework. On the gravity side, the missing mass problem \cite{zwicky33, rubin70},
and  the acceleration of the cosmic expansion \cite{riess98, perlmutter99} have motivated the search
for modifications of GR. On the  particle-physics side, the peculiar structure of the Standard Model (SM),}
the hierarchy of particle masses, the observed preponderance of matter over antimatter, the presence of
 several different gauge symmetries with a curious symmetry breaking pattern, are some of the puzzles that
motivate the search for extensions of the SM (notably supersymmetric ones \cite{fayet77}).

Most of the attempts to go beyond GR or beyond the SM, including the attempts to unify
all interactions, have suggested the existence of new particles and of new interactions.
In many cases, these new interactions give rise to apparent violations of the EP by predicting
additional long-range feeble forces that do not couple, as Einsteinian gravity does, to the total mass-energy
of a body. For instance, many theories including those with extra dimensions, from the Kaluza-Klein model \cite{Kaluza:1921tu,Klein:1926tv}
up to string theories \cite{Scherk:1974ca},
suggest the existence of a light spin-0, dilaton-like, particle. Such a light scalar field can be made compatible with 
current solar system tests if some screening mechanism is at work \cite{vainshtein72,damour94,Damour:1992kf,khoury04a, khoury04b, babichev09,hinterbichler10, brax13,burrage18}.
The coupling to matter of  a dilaton-like particle is expected to violate the EP at a small level 
(see, e.g., Refs. \cite{damour94, damour02,khoury04b,Damour:2010rp}). 
Another possibility is the existence of a very light spin-1 U boson,
related to an extension of the SM gauge group, mediating a new EP-violating force \cite{fayet90, fayet17}.}

{The EP, or more precisely the weak equivalence principle (WEP)} states that two bodies of different compositions and/or masses fall at the same rate in the same gravitational field (universality of free fall-UFF); equivalently, it states the equivalence of the ``inertial'' and ``gravitational'' masses. {Since its use by Einstein in 1907 as a starting point of GR, it has been experimentally tested with 
higher and higher precision.}  Tests of the WEP are usually presented in terms of the E\"otv\"os ratio $\eta$ \cite{eotvos22}, defined as the normalised difference of accelerations (or equivalently, as the normalised difference of gravitational-to-inertial {mass ratios}) of two test bodies {in} same gravitational field \cite{will14}:
\begin{equation} \label{eq_eotvos}
  \eta(2,1) = 2 \frac{a_2-a_1}{a_2+a_1} = 2 \frac{m_{G2}/m_{I2} - m_{G1}/m_{I1}}{m_{G2}/m_{I2} + m_{G1}/m_{I1}}
\end{equation}
where $a_j$ is the acceleration of the $j$th test-body, and $m_{Gj}$ and $m_{Ij}$ are its gravitational and inertial masses. {Since previous experiments have shown that $m_{G}/m_{I}$ does not differ from 1 by more than about $10^{-13}$, the quantity which we directly measure,}
\begin{equation}
\delta(2,1) = \frac{m_{G2}}{m_{I2}} - \frac{m_{G1}}{m_{I1}},
\end{equation}
{differs from the E\"otv\"os parameter $\eta(2,1)$ only by terms of order  $[\eta(2,1)]^2$}, and may in practice be identified with it.

{Tests of the UFF have} a long history, starting with Galileo Galilei (1638) and Newton (1687), and continuing to the end of the 20th century after Fischbach et al. \cite{fischbach86} revived the interest in experimental searches for new, WEP-violating interactions.
The state-of-the-art experiments have measured $|\eta|<{\rm a \,\, few\,\,} 10^{-13}$ (see Ref.~\cite{will14} for a historical account of tests of the WEP): (i) the E\"ot-Wash group used a high-precision torsion pendulum in the Earth and Sun gravitational fields \cite{schlamminger08, wagner12}, and (ii) {Lunar Laser Ranging has been used to monitor} the motions of the Moon and the Earth around the Sun \cite{williams12,viswanathan18}, {leading to  a slightly better accuracy; but in this case $\eta$ tests combination of effects due to composition differences between Earth and Moon (related to the WEP), and effects due to the self-gravity of each body (related to the strong EP).} 

{Concepts for an EP test in space were first developed in Stanford} (STEP project) to cope with {ground experiments limitations} \cite{chapman01, everitt2003}.
MICROSCOPE {was the first space experiment to test the WEP and hence also UFF.}
{Test masses in orbit follow  quasi-infinite and purer free falls in a quieter environment free of seismic disturbances and anthropogenic electromagnetic perturbations.}
The satellite was launched into a low-Earth, 710 km sun-synchronous orbit {by a Soyuz rocket} from Kourou on April 25, 2016. {It delivered data for more than two years.}

\begin{figure} 
  \center
  \includegraphics[width=0.45\textwidth]{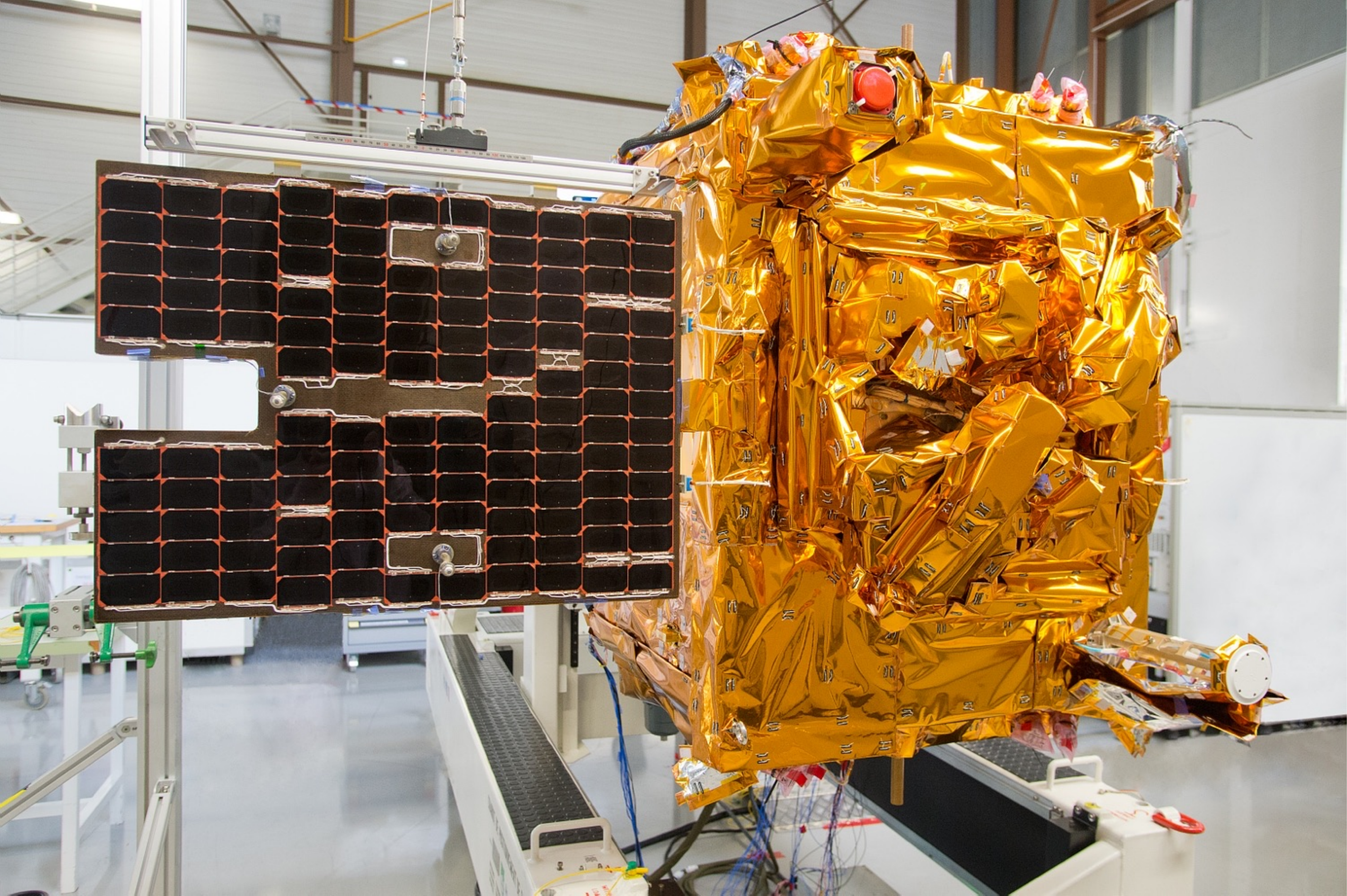}
  \includegraphics[width=0.35\textwidth]{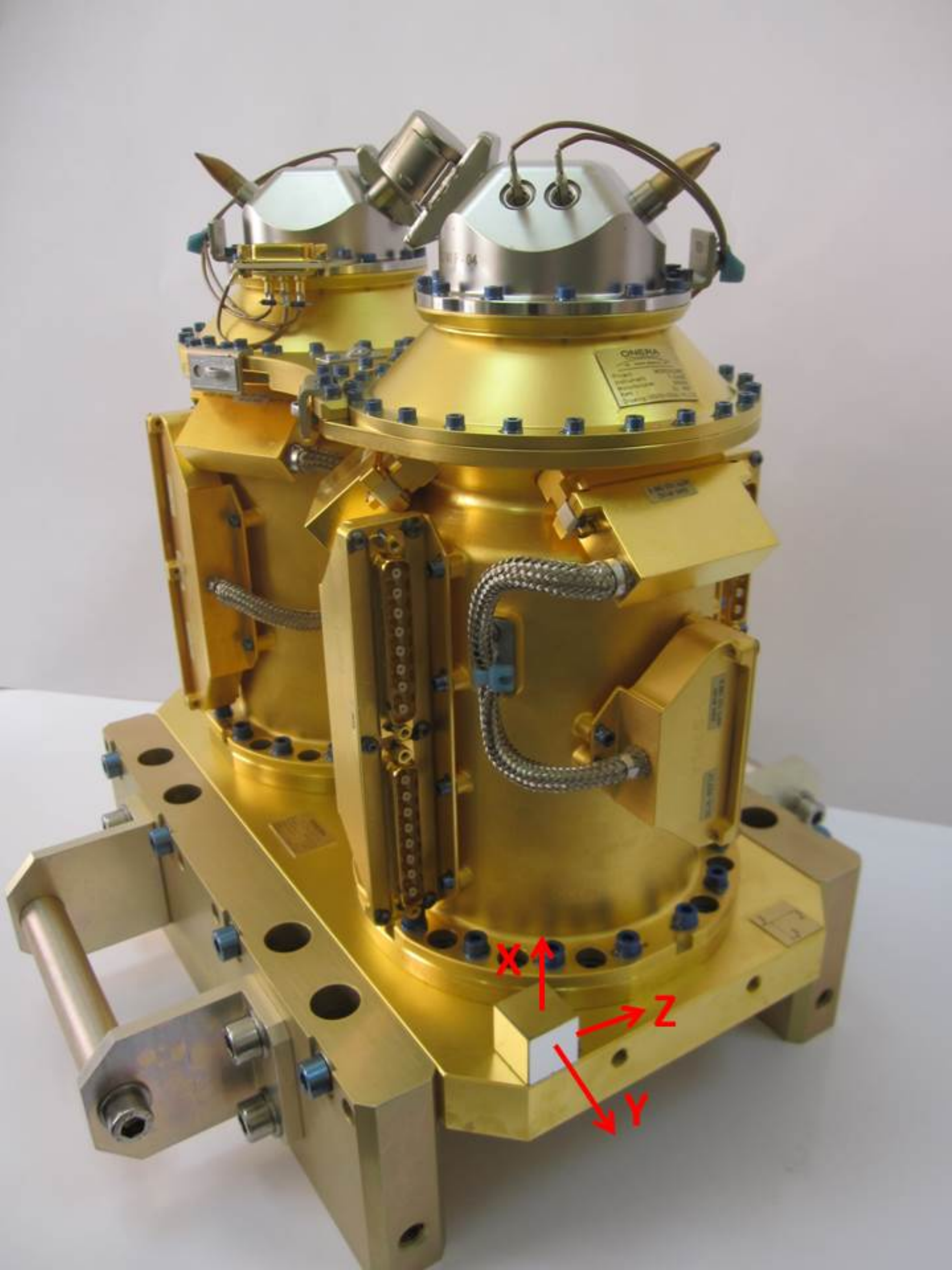}
\caption{The MICROSCOPE satellite{(left panel)} and the T-SAGE instrument with its two cylindrical sensor units {(right panel)}.}
\label{fig_TSAGE}       
\end{figure}

\begin{figure} 
  \center
  \includegraphics[width=0.40\textwidth]{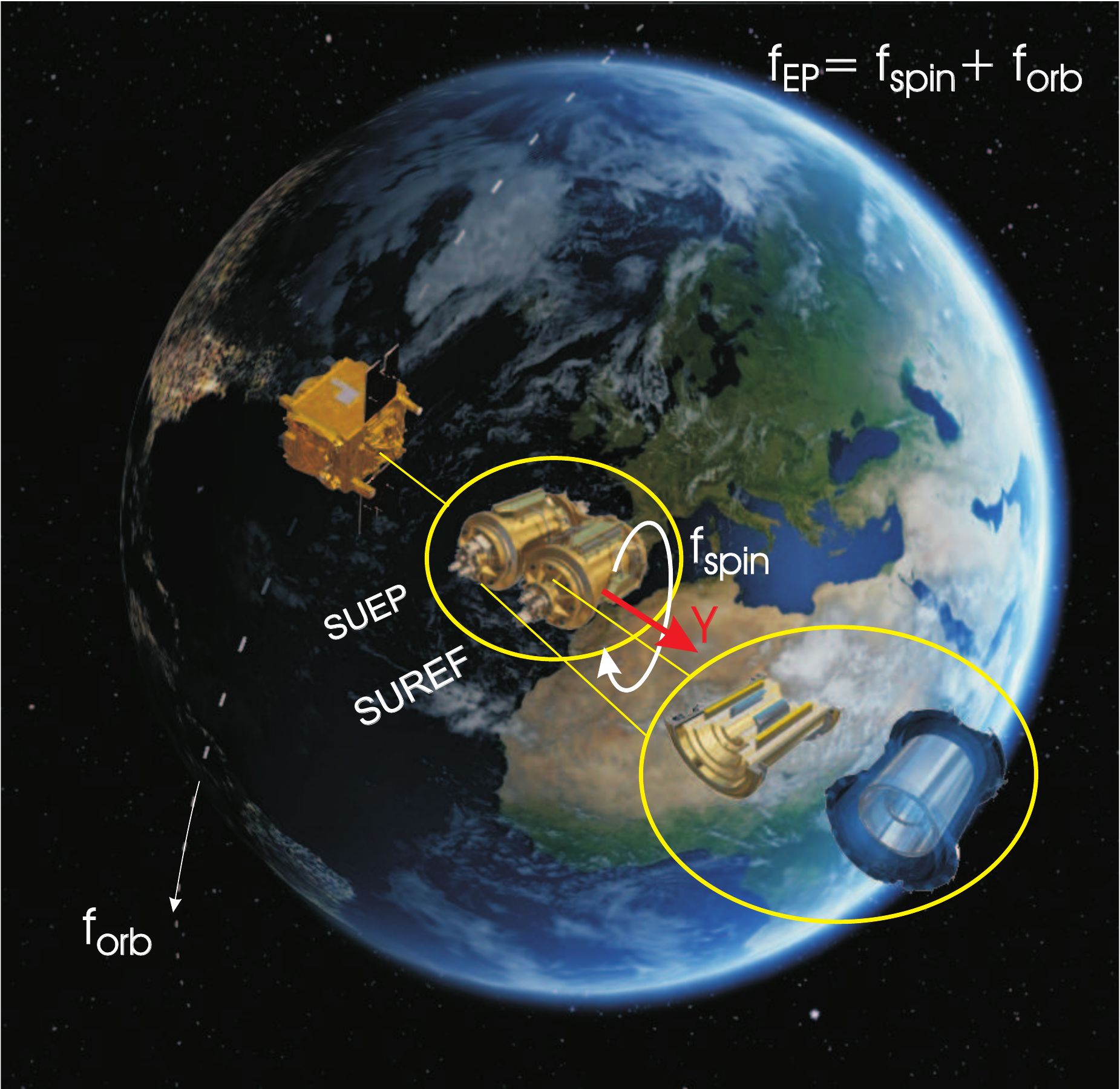}
  \includegraphics[width=0.40\textwidth]{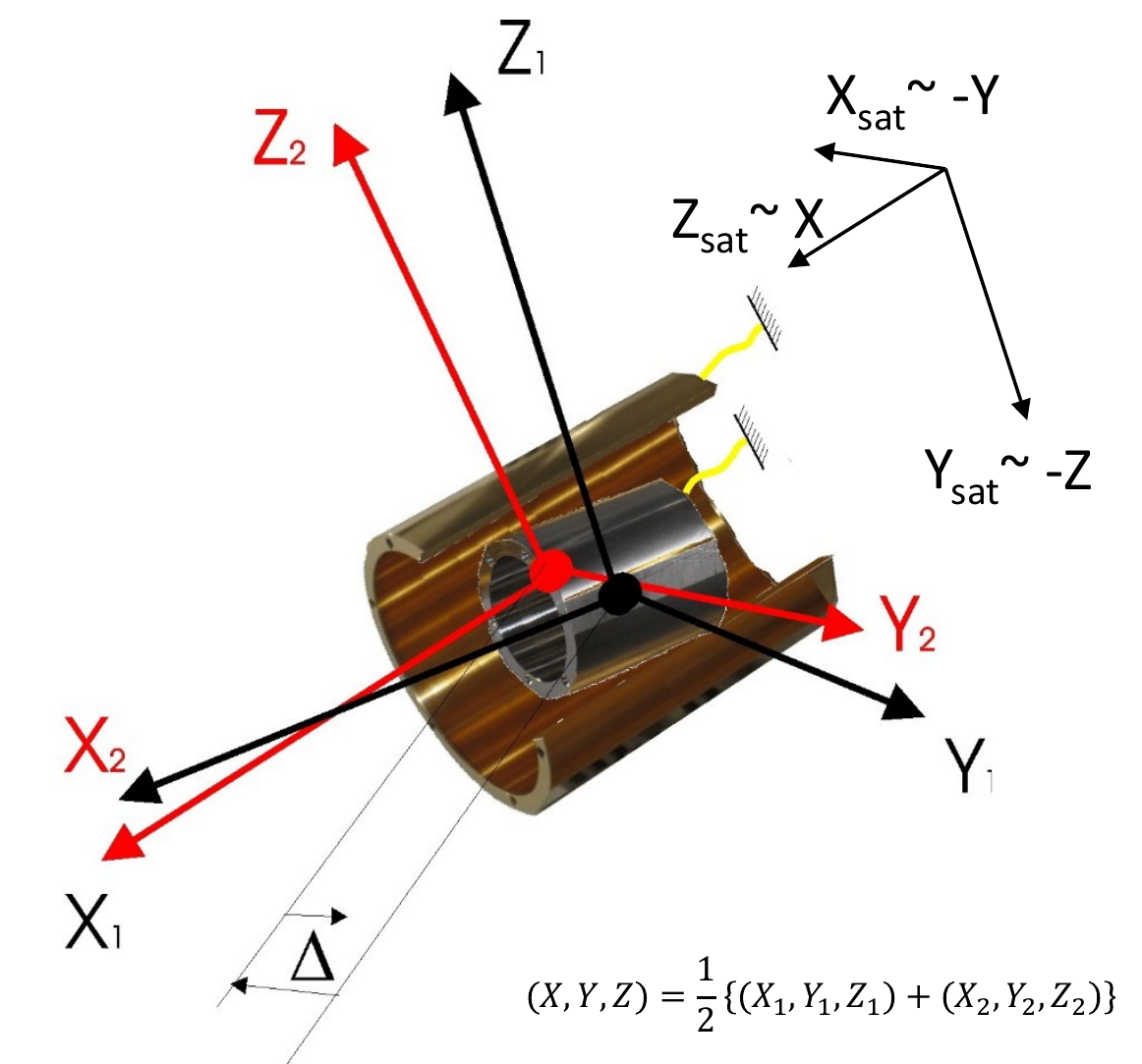}
\caption{Left: the  4 test-masses orbiting around the Earth (credits CNES / Virtual-IT 2017). Right: {reference} frames for the satellite and for one pair of masses. The ($X_{\rm sat}$, $Y_{\rm sat}$, $Z_{\rm sat}$) triad defines the satellite frame; the reference frames ($X_k$, $Y_k$, $Z_k$, $k=1,2$) are attached to the test-masses (black for the inner mass $k=1$, red for the outer mass $k=2$); the $X_k$ axes are the test-mass cylinders' longitudinal axis and define the direction of WEP test measurement; the $Y_k$ axes are normal to the orbital plane, and define the rotation axis when the satellite spins; the $Z_k$ axes complete the triads.  The 7 $\mu$m gold wires connecting the test-masses to the common Invar sole plate are shown as yellow lines. $\vec{\Delta}$ represents the test-masses offcentring. The centers of mass have been approximately identified with the origins of the corresponding sensor-cage-attached reference frames.}
\label{fig_config}       
\end{figure}

The satellite carries the Twin Space Accelerometers for Gravitation Experiment (T-SAGE) payload  (Fig.~\ref{fig_TSAGE}).
T-SAGE is composed of two sensor units called SUREF (Sensor Unit for Reference) and SUEP (Sensor Unit for the Equivalence Principle test). Each sensor unit includes two inertial sensors (or accelerometers), each one controlling one test mass. 
{The test masses are concentric, co-axial hollow cylinders. Choosing cylindrical shapes for the test masses allows (i) to nest them with their centres of mass at the same position, (ii) to approximate their tensor of inertia to that of a sphere with appropriate choice of dimensions\footnote{but unlike a sphere, the momenta of order $>2$ are not null.}, and (iii)  to optimise the capacitive sensing along the cylinders' common axis}\cite{liorzoucqg2}.
SUREF's test-masses are made of the same material (PtRh10), while SUEP's are made of different material (PtRh10) for the inner mass, Ti alloy for the outer mass  \cite{touboul19}).
{The PtRh10 platinum-rhodium alloy contains 90\% by mass of Pt (A = 195.1, Z = 78) and 10\% of Rh (A = 102.9, Z = 45). The isotopic composition
  of Pt has been measured by PTB on a sample of flight material\cite{touboul19}. SUEP’s outer test-mass is made of 90\% titanium (A = 47.9, Z = 22), 6\% of aluminium (A = 27.0, Z = 13) and 4\% of vanadium (A = 50.9, Z = 23).}
{The choice of the materials is a trade-off between machining laboratory know-how and theoretical motivation}\cite{touboul01b}. {Titanium and platinum differ mainly from the neutron excess over the atomic mass (N-Z)/A and, to a smaller extent, in the nuclear-electrostatic-energy parameter Z(Z-1)/(N+Z)$^{1/3}$.}
The instrument works by measuring the electrostatic force required to equilibrate all other “natural” forces in order to keep the test masses {motionless} with respect to electrodes fixed to the satellite \cite{liorzoucqg2}. The measured electrostatic force divided by the known mass is commonly called ``measured acceleration'' (this is the opposite of the acceleration which would be  undergone by the test mass in the absence of electrostatic force) {and this terminology will be used in this paper.} The driving idea of the experiment is to compare the measured accelerations {of the two test mass pairs within each sensor} to verify if they have the same free-falls.
{Along the $X$ axis, parallel to the cylinder axis} (Fig.\ref{fig_TSAGE}, right panel and Fig.~\ref{fig_config}, right panel), {capacitance changes are caused by the variation of the overlap between the test mass and its surrounding electrodes fixed to the satellite.  Along the $Y$ and $Z$ axes the capacitance changes through variation of the gap. This allows for better sensitivity, removes electrostatic instability, and gives complete linearity along $X$. This is why the $X$-axis will be used in our analyses.}

The MICROSCOPE satellite was designed to provide an environment as stable as possible. It was finely controlled along its six degrees of freedom {with a} Drag-Free and Attitude Control System (DFACS) described in \cite{robertcqg3}. The DFACS allows several modes of operation: inertial pointing or spin mode with the choice between several rotation rates. In all cases the instrument's $X$-axis is kept parallel to the orbital plane (Fig.~\ref{fig_config}, {left panel).} {In inertial pointing mode the axes of the spacecraft and instrument are  maintained pointing in a fixed direction and hence the direction of the Earth gravity field projected onto the $X$-axis varies at the orbital frequency and hence $f_{\rm{EP}}=f_{\rm{orb}}$. In the spin mode the satellite is rotated about the instrument $Y$-axis, which is orthogonal to the orbital plane, at a frequency  $f_{\rm{spin}}$, and in this case  $f_{\rm{EP}}=f_{\rm{orb}}+f_{\rm{spin}}$.} The measurements analysed in this paper were obtained with two different spin rates referred to as $f_{\rm{spin}_2}$ and $f_{\rm{spin}_3}$. The values of the frequencies of interest in this paper are given in Table \ref{tab_freq} according to \cite{rodriguescqg1}. {The higher frequency, $f_{\rm{spin}_3}$, is a trade-off between minimisation of instrument noise, which would favour a higher frequency, and the capabilities of the micro-propulsion system; the smaller value, $f_{\rm{spin}_2}$, has the advantage of being farther from the limits of the propulsion system, and saves on gas usage.}

\begin{table}
\caption{\label{tab_freq} Main frequencies of interest.}
\begin{indented}
\item[]\begin{tabular}{@{}llll}
\br
Label & Frequency & Comment \\
\mr
$f_{\rm{orb}}$ & $0.16818\times{}10^{-3}$ Hz & Mean orbital frequency \\
\mr
$f_{\rm{spin}_2}$ & $\frac{9}{2}f_{\rm{orb}}$=$0.75681 \times{}10^{-3}$ Hz& Spin rate frequency 2 (V2 mode) \\
\mr
$f_{\rm{spin}_3}$ & $\frac{35}{2} f_{\rm{orb}}$=$2.94315 \times{}10^{-3}$ Hz& Spin rate frequency 3 (V3 mode) \\
\mr
$f_{\rm{EP}_2}$ & $0.92499 \times{}10^{-3}$ Hz& EP frequency in V2 mode  \\
\mr
$f_{\rm{EP}_3}$ & $3.11133 \times{}10^{-3}$ Hz& EP frequency in V3 mode \\
\br
\end{tabular}
\end{indented}
\end{table}

In \cite{touboul17}, {we used} 7\% of the available data to provide first results. 
No evidence for a violation of the EP was found at the $1.3\times{}10^{-14}$ level in terms of the E\"otv\"os parameter, {also providing improved constraints on additional new long-range forces \cite{berge18, fayet18, fayet19}}.
Since then, the use of all data has allowed us to improve significantly the statistical error, and a thorough analysis of systematic errors has been conducted \cite{hardycqg6} {using} additional calibration sessions. {The present} paper describes these efforts and their result.  
In Sect.~2, we review the available data at our disposal, grouped into scientific sessions.  In Sect.~3, we explain the physical parametric model used to analyse the measurements. In Sect.~4 we show that perturbed behaviours {of the measured accelerations} lasting one or several orbits {requires the division of} some sessions into several disjointed segments. We list the characteristics of the segments analysed in this  paper and describe the glitch (short singular events) detection strategy. The analysis is performed first on {individual} segments as shown in Sect.~5, and then using the  data from all segments gathered for a {single} estimation as presented in Sect.~6. We conclude in Sect.~7.

MICROSCOPE is simple in its principle but each component {of the mission} has been pushed to its limits given the external constraints (e.g. size of the satellite and global cost); {for a more detailed presentation of MICROSCOPE the reader is referred to} other papers of this volume (Ref.~\cite{touboulcqg0, rodriguescqg1, liorzoucqg2, robertcqg3, rodriguescqg4, chhuncqg5, hardycqg6, bergecqg7, bergecqg8}).

\section{Scientific sessions and available data}
\label{sec:scient-sess-avail}

The MICROSCOPE {observations are} divided {into} different measurement sessions. {A session represents} a time span during which the satellite and the instrument keep the same configuration (spin, drag-free {mode, etc}). They are numbered by increasing integers and are described in the mission scenario \cite{rodriguescqg4}.  Some of these sessions  (called ``EP sessions'') are directly devoted to the test of the Equivalence Principle while others (``calibration sessions'') are used to calibrate or characterise the experimental apparatus \cite{chhuncqg5}. EP sessions  are the longest, most of them {spanning} 120 orbital periods (about 8 days), while calibration sessions typically {span} a few orbits. All sessions {were} performed with SUEP as well as with SUREF. Sessions are characterised by:
\begin{itemize}
\item The sensor unit (SU) {used}: as explained in \cite{liorzoucqg2}, the payload is composed of 2 SUs, each SU enclosing 2 {co-axial} concentric test masses. SUREF has 2 test masses with the same composition and is used {as a null check of the experiment}; SUEP aims at comparing the free falls of a test mass in platinum and of a test mass in titanium. During most of the sessions  only one SU is on, {with both} operating simultaneously only during the EP sessions 430, 452 and 454.
\item {Which combination of the accelerometer outputs are used by the DFACS}\cite{robertcqg3}: the DFACS  uses micro-thrusters servo-controlled by the outputs of one or several inertial sensors in order to cancel their measured acceleration and to stabilise the rotation;  it can be controlled by the output of one of the two {inertial sensors (labeled IS1 for the inner mass and IS2 for the outer mass) within the SU in operation.  A common mode  use of both accelerometers is also possible.} In practice almost all sessions used IS2 except sessions 358 and 406 which used the common mode. When both the SUs are working, the DFACS is controlled {by the accelerometers within only one of the SUs}: SUEP for session 430 and SUREF for sessions 452 and 454. 
\item The spin rate of the satellite: either V2 corresponding  to the frequency $f_{\rm spin_2}=  (9/2) f_{\rm orb} \simeq 7.57\times{}10^{-4}$ Hz  or V3 corresponding  to  $f_{\rm spin_3}= (35/2) f_{\rm orb} \simeq 2.94\times{}10^{-3}$ Hz.
\item {The session duration: sessions were planned to last as long as possible, subject to operational constraints}\cite{rodriguescqg4}: {(i) periodic pointing updates required to remain in specification due to possible onboard clock drift; this limited the maximum duration to 120 orbits, each of duration $T_{\rm orb}=5946$; (ii) roughly once a month the star trackers pointed towards the bright moon and the fine attitude control had to be interrupted so that some sessions had to be shortened. In addition a few sessions were interrupted due to technical problems.}
\end{itemize}
The first EP session after the commissioning is session 120, performed with SUREF. Note, the first calibration sessions revealed significant non-linearities in SUEP, which were solved by modifying {the parameters of} the proportional–integral–derivative (PID) controller  in the servo-loop of  SUEP, and its behaviour was nominal from session 210 onwards. {None of the prior sessions are used in our analysis.} Session 430 {was interrupted by an anomaly and was discarded}. Finally we are left with 9 EP sessions performed {with} SUREF (Table \ref{tab-sessions-SUREF}) and 18 {with} SUEP (Table \ref{tab-sessions-SUEP}). 

Tables \ref{tab-sessions-SUREF} and  \ref{tab-sessions-SUEP} list also {the sensors' minimum and maximum} temperatures. The typical temperature {of} the SUEP was about 10$^\circ$C while it was about 19$^\circ$C {for}  the SUREF. The higher temperature in SUREF {was due to two defective capacitors} used for house-keeping data  \cite{liorzoucqg2}. Note also the higher temperatures during sessions 452 and 454 when the two SUs ran simultaneously.

\begin{table}
  \caption{\label{tab-sessions-SUREF}List of analysed sessions dedicated to the EP test with SUREF and their characteristics. {Column 3 gives the calendar date of the beginning of each session, whereas columns 4 and 5 give the beginning and the end in terms of orbit number. The counting starts on 2016-04-26T01:03:05 UTC  (orbit 1) and the orbital period is 5946 s.} Columns 7 and 8 give the minimal and maximal temperatures {measured in the sensor unit} during the session.}

\begin{indented}
\lineup
\item[]\begin{tabular}{@{}lllrrrrr}
\br                              
$\0\0$Session&Spin&Beginning&{Beginning}      & {End\ \ \ }               &Duration &Tmin          & Tmax\cr 
$\0\0$number&       &  (date)    &(orbit N$^{\circ}$)&(orbit N$^{\circ}$)&(orbits)   &($^\circ$C) &  ($^\circ$C) \cr 
\mr
\0\0120&V2&2016-10-27&2681.5&2801.5&120.0 &18.1  &18.4  \cr
\0\0174&V2&2017-01-19&3952.8&4072.8&120.0 &18.1  &19.0  \cr
\0\0176&V2&2017-01-27&4074.4&4156.4&82.0   &19.0  &19.0  \cr
\0\0294&V3&2017-09-13&7397.6&7491.6&94.0   &17.5  &18.3  \cr
\0\0376&V2&2017-11-11&8258.0&8338.0&80.0   &16.1  &18.0  \cr
\0\0380&V3&2017-11-20&8382.6&8502.6&120.0 &19.2  &19.4  \cr
\0\0452&V2&2018-02-05&9507.7&9541.7&34.0   &34.2  &35.3  \cr
\0\0454&V2&2018-02-07&9542.6&9607.8&65.2   &35.3  &35.7  \cr
\0\0778&V2&2018-09-28&12930.0&12990.01&60.0   &20.0  &21.3  \cr
\br
\end{tabular}
\end{indented}
\end{table}

\begin{table}
  \caption{\label{tab-sessions-SUEP}Same as Table \ref{tab-sessions-SUREF} for sessions performed {with} SUEP.} 
  
\begin{indented}
\lineup
\item[]\begin{tabular}{@{}lllrrrrr}
\br                              
$\0\0$Session&Spin&Beginning&{Beginning}      & {End\ \ \ }               &Duration &Tmin          & Tmax\cr 
$\0\0$number&       &  (date)    &(orbit N$^{\circ}$)&(orbit N$^{\circ}$)&(orbits)   &($^\circ$C) &  ($^\circ$C) \cr 
\mr
\0\0210&V3&2017-02-14&4336.5&4386.5&50.0    &9.7  & 10.4  \cr
\0\0212&V3&2017-02-18&4388.0&4464.1&76.1 &10.4&10.4   \cr
\0\0218&V3&2017-02-28&4535.1&4655.1&120.0  &9.1  &10.3   \cr
\0\0234&V3&2017-03-15&4751.1&4843.1&92.0    &9.3  &10.3   \cr
\0\0236&V3&2017-03-21&4844.6&4964.6&120.0  &10.3&10.4   \cr
\0\0238&V3&2017-03-29&4966.1&5086.1&120.0  &10.4&10.6   \cr
\0\0252&V3&2017-04-13&5176.7&5282.7&106.0  &9.2  &10.7   \cr
\0\0254&V3&2017-04-20&5284.3&5404.3&120.0  &10.7&10.8   \cr
\0\0256&V3&2017-04-29&5405.8&5525.8&120.0  &10.8&10.8   \cr
\0\0326&V3&2017-09-27&7600.5&7702.5&102.0  &10.1&12.9   \cr
\0\0358&V3&2017-10-14&7857.3&7950.1&92.8 &9.4  &9.9   \cr
\0\0402&V2&2017-12-06&8614.7&8634.7&20.0    &10.0&10.3   \cr
\0\0404&V3&2017-12-07&8637.8&8757.8&120.0  &10.3&11.7   \cr
\0\0406&V3&2017-12-16&8759.3&8779.3&20.0    &11.7&11.7   \cr
\0\0438&V2&2018-01-16&9215.2&9255.2&40.0    &9.8  &10.7   \cr
\0\0442&V2&2018-01-22&9298.3&9338.3&40.0    &11.2&11.4   \cr
\0\0748&V2&2018-09-03&12562.3&12586.3&24.0    &9.3  &9.7   \cr
\0\0750&V3&2018-09-05&12589.3&12597.3&8.0      &9.7  &9.8   \cr
\br
\end{tabular}
\end{indented}
\end{table}

All sessions come with the following data  used directly to estimate the E\"otv\"os parameter \cite{bergecqg7}:
\begin{itemize}
\item The {measured} accelerations (with the meaning explained in section 1) for each test mass of the operating SU at a sampling rate of 4 Hz with the associated {time stamping}. The difference of acceleration $\Gamma_x^{(d)} $ between the 2 test masses along the most sensitive axis $X$ (see \cite{rodriguescqg1} and \cite{liorzoucqg2} for the description of the axis), {which was analysed in this work}, is directly computed from these data.
\item The attitude, angular velocity and angular acceleration of the satellite with respect to the inertial reference frame J2000 \cite{robertcqg3}. These data are given at exactly the same {time stamps} as the accelerometer measurements.
\item The position and velocity of the centre of mass of the satellite in the J2000 frame sampled every minute.
\end{itemize}
Additionally, housekeeping data (sampled at 1 Hz) are used to monitor the behaviour of the experiment and to estimate systematic errors \cite{chhuncqg5, hardycqg6}:
\begin{itemize}
\item The variation of position of each test-mass as measured by the capacitive sensors; the residual displacements are very small (less than $10^{-12}$ m at $f_{\rm EP}$)  and the corresponding acceleration is negligible compared to our needs (Sect. \ref{sec:impact-resid-vari}).
\item The temperature {which is} measured by several probes inside the mechanical {and electrical subsystems of each SU}\cite{liorzoucqg2}. The corresponding systematic effects are estimated in \cite{hardycqg6}.
\end{itemize}

\section{The measurement model}
\label{sec:measurement-model}
{A detailed explanation of the measurement model is given in} \cite{rodriguescqg1}. {Here we only summarise its main aspects.}
As explained in the introduction, we look for a difference of free fall between two {co-axial} concentric test masses of the same SU (SUREF or SUEP) by analysing the difference of their measured accelerations $\vv\Gamma^{(d)} =\vv\Gamma^{(1)} -\vv\Gamma^{(2)} $. In a perfect experiment, this would correspond exactly to the applied differential acceleration  $\vv\gamma^{(d)} =\vv\gamma^{(1)} -\vv\gamma^{(2)} $. Additional terms must be considered to account for the real experiment: 
\begin{itemize}
\item A mapping matrix $\left[ {\mathbf {\tilde{A}}^{(c)}}\right] $\footnote[1]{{In the whole paper the notation $\left[ {\mathbf {M}}\right] $ designates a matrix.}} between the applied and the measured acceleration; { $\left[ {\mathbf {\tilde{A}}^{(c)}}\right] $ is close to the identity matrix and} takes into account scale factors and coupling between axes as well as a rotation reflecting the fact that our model for $\vv\gamma^{(d)}$ is not exactly expressed in the instrument frame which is imperfectly known:  $\vv\gamma^{(d)}\ \longrightarrow\ \left[ {\mathbf {\tilde{A}}^{(c)}}\right] \vv {\gamma}^{(d)}  $; 
\item A residual projection of the measured  common mode acceleration $2\left[ {\mathbf {\tilde{A}}^{(d)}}\right]  \vv {\tilde{\Gamma}}^{(c)} $  where $ \vv {\tilde{\Gamma}}^{(c)}=\left(\vv {\tilde{\Gamma}}^{(1)}+\vv {\tilde{\Gamma}}^{(2)} \right)/2$ and  $\vert\vert\left[ {\mathbf {\tilde{A}}^{(d)}}\right] \vert\vert \ll{}1$; formally, $\vv {\tilde{\Gamma}}^{(i)}$ are the noise-free measured accelerations but in practice, we approximate them with  $\vv\Gamma^{(i)}$; 
    \item A coupling with the angular acceleration $\dt{\vv \Omega} $ (because the sensors measure both linear and angular accelerations): $2\left[{\mathbf C'^{(d)}} \right]  \dt{\vv \Omega}$;
  \item  A measurement bias $\vv{B_0}^{(d)}$;
  \item A noise $ \vv n^{(d)}$.
  \end{itemize}
  The origin of these terms is related  both to detailed instrumental characteristics \cite{liorzoucqg2} and to the implementation of the instrument in its environment \cite{rodriguescqg1}.
We end up with the model
\begin{equation} 
\vv\Gamma^{(d)} = \vv{B_0}^{(d)}+ \left[ {\mathbf {\tilde{A}}^{(c)}}\right] \vv {\gamma}^{(d)} +2\left[ {\mathbf {\tilde{A}}^{(d)}}\right]  \vv {\tilde{\Gamma}}^{(c)}+2\left[{\mathbf C'^{(d)}} \right]  \dt{\vv \Omega} + \vv n^{(d)}.
\label{eq_diffT}
 \end{equation} 
{Note that this equation is fully similar to Eq. (19) of} Ref.~\cite{rodriguescqg1} {but we have used the notation shortcut $\left[ {\mathbf {\tilde{A}}^{(d)}}\right]=\left[{\mathbf A^{(d)}} \right] \left[{\mathbf A^{(c)}} \right]^{-1} $.} The derivation of this model and in particular  how we get a mixing between the applied differential acceleration and the measured common mode acceleration is detailed in  \cite{rodriguescqg1}. Note also that other potential disturbing effects (non-linearity, thermal effects, stiffnesses and others) are not included in this model but are characterised in \cite{chhuncqg5} and \cite{hardycqg6} and  contribute to the assessment of the systematic effects \cite{hardycqg6}.

 The difference of accelerations applied to the test masses derives directly from simple dynamics.
 First, each mass {experiences} Earth gravity; the gravitational force applied to each mass is slightly different because of (i) a gravity gradient due to their small difference of positions $\vv{\Delta} $ {(called offcentring in the following, right panel of} Fig.~\ref{fig_config}), and (ii) of a possible intrinsic difference of free fall parametrised by the E\"otv\"os parameter. 
 Second, since the acceleration is expressed in the instrument frame co-rotating with the satellite at an angular velocity matrix $[\Omega]$, the corresponding inertial acceleration must be taken into account.
 Third, additional small perturbations on the test masses (local gravity, magnetic effects, radiation pressure, radiometric effect and others)  are gathered in the {physical} bias $\vv{b_1}^{(d)}$. The detailed derivation \cite{rodriguescqg1} yields 
\begin{equation} \label{eq_gamma}
\vv\gamma^{(d)} = \delta(2,1) \vv{g}(O_{\rm sat}) + ([{\rm T}] - [{\rm In}]) \vv{\Delta} + \vv{b_1}^{(d)},
\end{equation}
where
\begin{itemize}
\item $ \delta(2,1)=m_{\rm{G_2}}/m_{\rm{I_2}} - m_{\rm{G_1}}/m_{\rm{I_1}}\simeq \eta({\rm{2, 1}})=-\eta({\rm{1, 2}}$) {(note that Eq.~(\ref{eq_gamma}) involves $\delta(2,1)$ instead of $\delta(1,2)$ because  the measured differential acceleration is opposite to the difference of gravity accelerations),}
\item $\vv{g}(O_{\rm sat}) $ is the gravity acceleration computed at the centre of the satellite,
\item $[{\rm T}]$ is the gravity gradient tensor computed at the centre of the satellite,
\item $ [{\rm In}]=[\Omega]^2+[\dot\Omega]$, is the gradient of inertia matrix.
\end{itemize}
In this analysis, we will use only the measurement along the $X$-axis which is an order of magnitude more sensitive than {those along} $Y$ and $Z$. This leads to

\begin{equation}  \label{eq_xacc}
\begin{split}
\Gamma_x^{(d)} &\approx  B_{0x}^{(d)} \\
    &+  \tilde{a}_{c11} b_{1x}^{(d)} + \tilde{a}_{c12} b_{1y}^{(d)}+ \tilde{a}_{c13} b_{1z}^{(d)} \\
    &+ \tilde{a}_{c11} \delta g_x + \tilde{a}_{c12} \delta g_y + \tilde{a}_{c13} \delta g_z \\
    & + \left(T_{xx} - {\rm In}_{xx} \right) \tilde{a}_{c11} \Delta_x + \left(T_{xy} - {\rm In}_{xy} \right) \tilde{a}_{c11} \Delta_y + \left(T_{xz} - {\rm In}_{xz} \right) \tilde{a}_{c11} \Delta_z \\
    & +  \left(T_{yx} - {\rm In}_{yx} \right) \tilde{a}_{c12} \Delta_x + \left(T_{yy} - {\rm In}_{yy} \right) \tilde{a}_{c12} \Delta_y + \left(T_{yz} - {\rm In}_{yz} \right) \tilde{a}_{c12} \Delta_z \\
    & + \left(T_{zx} - {\rm In}_{zx} \right) \tilde{a}_{c13} \Delta_x + \left(T_{zy} - {\rm In}_{zy} \right) \tilde{a}_{c13} \Delta_y  + \left(T_{zz} - {\rm In}_{zz} \right) \tilde{a}_{c13} \Delta_z \\
    &+ 2\left( {\tilde{a}_{d11}} {\Gamma}_x^{(c)} + {\tilde{a}_{d12}}{\Gamma}_y^{(c)} +{\tilde{a}_{d13}} {\Gamma}_z^{(c)}  \right)\\ 
    &+ 2 \left(c'_{d11} \dt{\Omega}_x + c'_{d12} \dt{\Omega}_y + c'_{d13} \dt{\Omega}_z  \right)\\
    &+ n_x^{(d)}-2\left( {\tilde{a}_{d11}} n_x^{(c)} + {\tilde{a}_{d12}}n_y^{(c)} +{\tilde{a}_{d13}}  n_z^{(c)}  \right) ,\\
\end{split}
\end{equation}
where  {$\tilde{a}_{cij}$, $\tilde{a}_{dij}$, $c'_{dij}$ are the elements of  $ \left[ {\mathbf {\tilde{A}}^{(c)}}\right] $, $\left[ {\mathbf {\tilde{A}}^{(d)}}\right]$ and $\left[ {\mathbf C'^{(d)}}\right] $.}

As explained in \cite{bergecqg7}, some terms have been demonstrated to be negligible and others are corrected. Finally, after calibration and correction \cite{bergecqg7} { (and see also} Sect.~\ref{sec:main-steps-analysis} {for a summary),} we get the fundamental equation which will be used for our analysis \cite{bergecqg7}:
\begin{equation}
  \label{eq:11}
  \Gamma^{(d)}_{x, {\rm corr}}=\tilde{b}_x^{'(d)}+\delta_x g_x+\delta_z  g_z+\Delta'_{x}  S_{xx} +\Delta'_{z}  S_{xz}+ n_x^{(d)},
\end{equation}
where
\begin{itemize}
\item $\tilde{b}_x^{'(d)}$ is a bias {which is almost constant but may slowly drift over time due to thermal effects};
\item  $\left[{\mathbf S}\right]$ is the symmetric part  of  the  $\left[{\mathbf T}\right] - \left[{\bf In}\right]$ matrix;
\item $\delta_x=\tilde{a}_{c11} \delta \simeq \delta$ is very close to the E\"otv\"os parameter {since $\vert\tilde{a}_{c11}-1\vert <2\times{}10^{-2}$} whereas the potential contribution of the E\"otv\"os parameter to $\delta_z=\tilde{a}_{c13} \delta$ should be much smaller because $\vert \tilde{a}_{c13}\vert <2.6\times{}10^{-3}$ rad from manufacturing;
\item $\Delta'_{x}$ and $\Delta'_{z}$ are effective combinations of the components of the offcentrings between the 2 test masses \cite{bergecqg7}.
\end{itemize}
The structure of this equation is very simple:
\begin{itemize}
\item $g_x$, $g_z$, $ S_{xx}$ and  $ S_{xz}$ are time-varying deterministic signals which can be computed accurately \cite{bergecqg7} knowing the position and the pointing of the satellite as well as its angular velocity and acceleration which are all delivered by CNES \cite{rodriguescqg1} with an accuracy better than the requirements \cite{robertcqg3};
\item  $\tilde{b}_x^{'(d)}$ is taken into account by estimating a polynomial trend;
\item The parameters $\delta_x$, $\delta_z$, $\Delta'_{x}$ and $\Delta'_{z}$ are estimated.
\end{itemize}
These quantities are computed in the instrument frame, in which the varying signals of Eq.~(4) have very different frequency patterns \cite{touboul01b}:
\begin{itemize}
\item  $g_x$ and $g_z$ are essentially periodic signals of frequency $f_{\rm EP}$ and are in phase quadrature,
\item $ S_{xx}$ and  $ S_{xz}$ have dominant components at DC and $2 f_{\rm EP}$ and the variations of $ S_{xx}$ and  $ S_{xz}$ at $2 f_{\rm EP}$ are  in phase quadrature,
\item  $\tilde{b}_x^{'(d)}$ is at very low frequency.
\end{itemize}
As a consequence, these signals are almost uncorrelated. 

The bias, which encompasses the low frequency trend, is modelled with a degree 3 polynomial : $\tilde{b}_x^{'(d)}(t)=\sum_{j=0}^3\alpha_j(t-t_0)^j$. Substituting it in Eq.~\eqref{eq:11}, we finally get
\begin{equation}
  \label{eq:2}
   \Gamma^{(d)}_{x, {\rm corr}}=\sum_{j=0}^3\alpha_j(t-t_0)^j+\delta_x g_x+\delta_z  g_z+\Delta'_{x}  S_{xx} +\Delta'_{z}  S_{xz}+ n_x^{(d)}.
\end{equation}

\section{Handling of singular events}
\label{sec:handl-sing-events}

\subsection{Segmentation}
\label{sec:segmentation}

\begin{figure} 
\center
\includegraphics[width=0.45\textwidth]{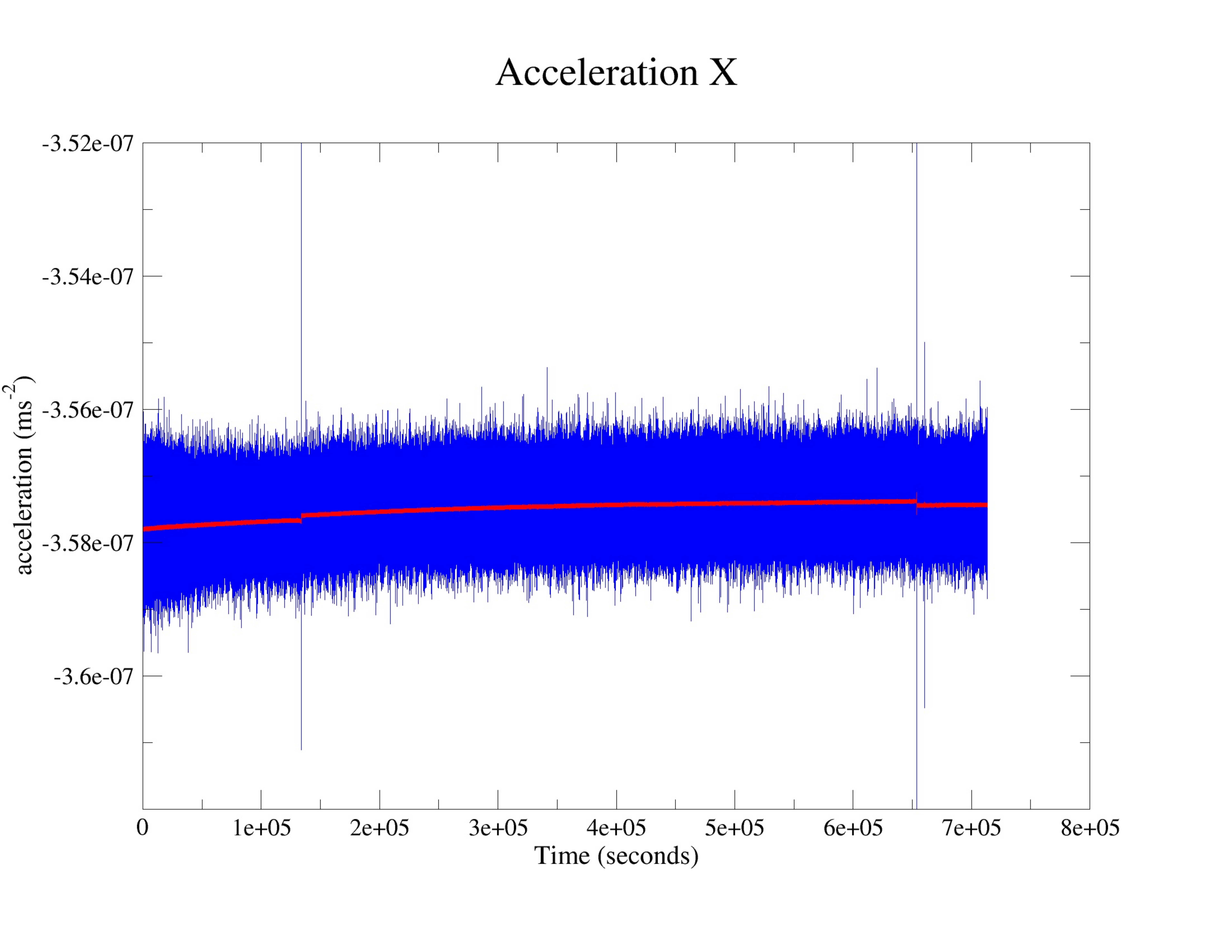}
\includegraphics[width=0.45\textwidth]{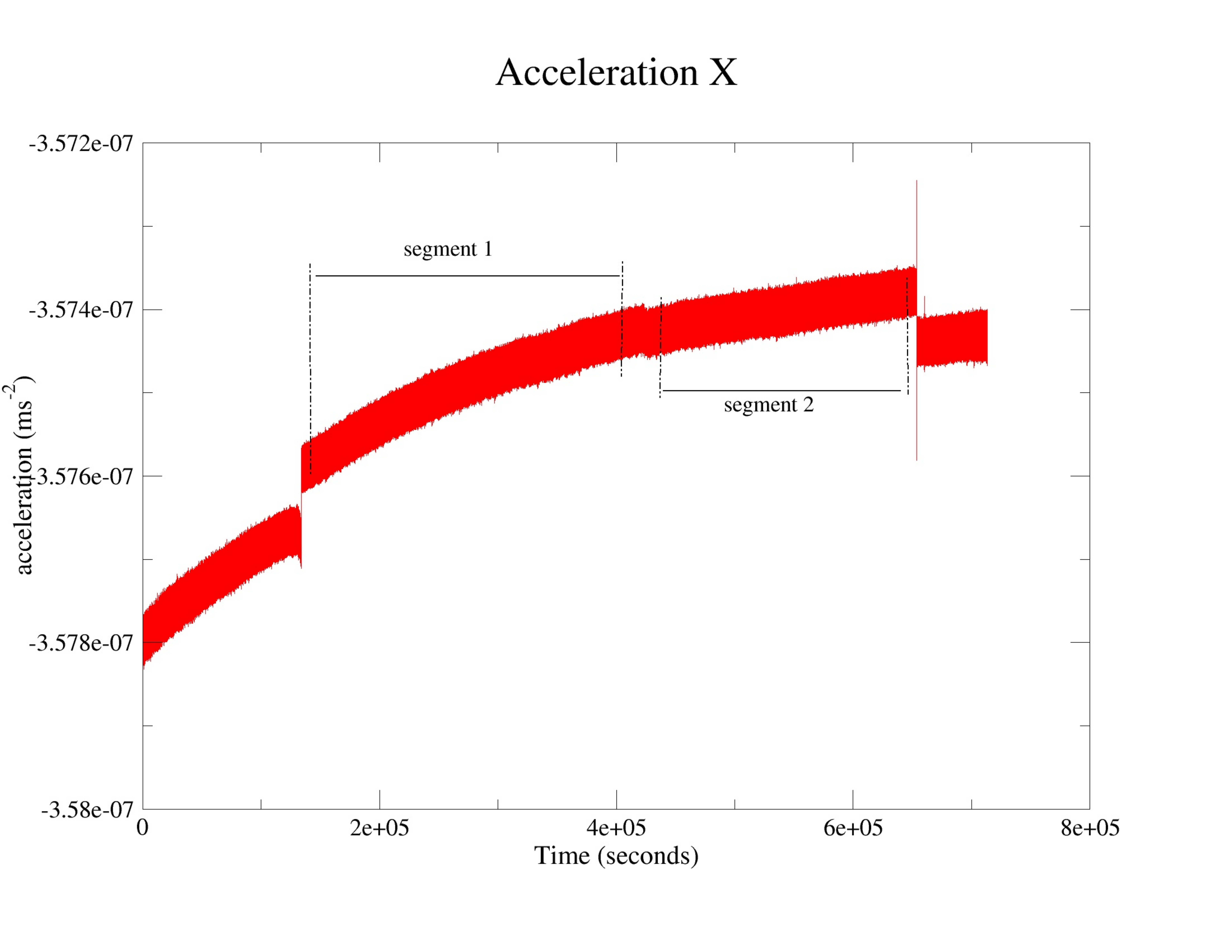}
\includegraphics[width=0.45\textwidth]{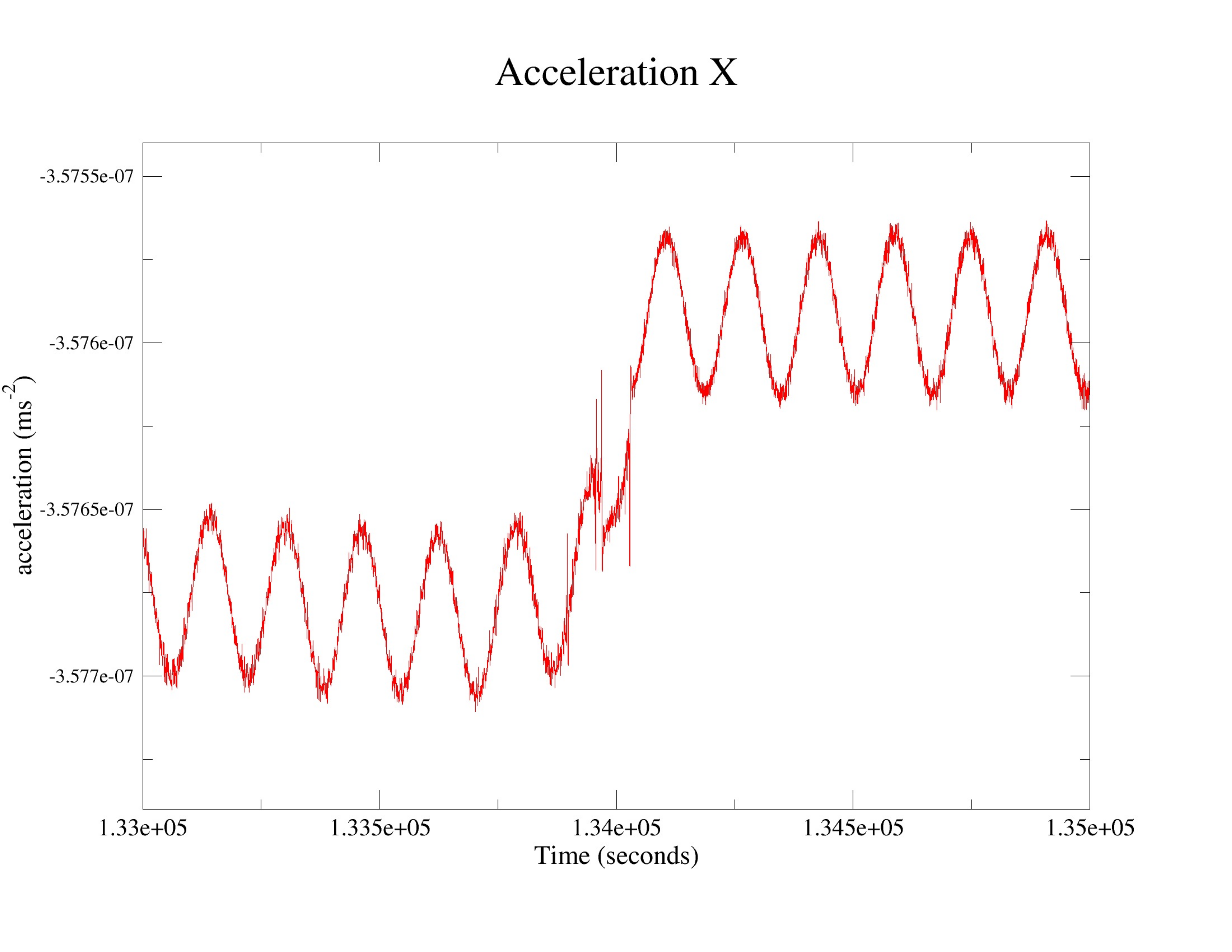}
\caption{Differential acceleration for session 380. The first graph (top left) shows raw data in blue and filtered (using a running average over 60 s) data in red. The red curve {reveals} leaps which are not clearly visible on raw data: this appears more clearly by zooming in the ordinate axis as shown on the second graph (top right).Finally the third graph (bottom) which zooms in on the first leap shows that this is not a simple step but that there are disturbed measurements in the neighbourhood. Consequently, in the {actual} analysis, we use only data belonging to the 2 segments represented {in} the top right panel.}
\label{fig_session380}       
\end{figure}

During some EP sessions, sudden changes in the local mean of the measured acceleration can be noted (Fig.~\ref{fig_session380}).  The typical macroscopic manifestation {in} the raw data (first panel of the figure) is a leap in the values of the measurements.
These leaps are observed on the SUREF instrument {only}. They are not well understood as they are unpredictable, rare and not correlated  to other observable events.
Applying a lowpass filter and then zooming on the leap  (third panel) shows that it does not consist {of} a simple Heaviside step function but {is} a complex mixing of erratic oscillations. {We are faced with at least a few hundred seconds of unusable data.}
More rarely  some sessions have been stopped after the detection of technical problems (for example an instability in the drag-free loop) and there was a delay between the occurrence of the problem and the {interruption}. {The data associated with the occurrence of such problems were discarded.}

Thus, some sessions cannot be analysed in full, but only in parts {referred to as "segments" in the following}. There can be a single segment if only the end of the session is corrupted or if the other potential usable parts are too short to bring a significant contribution; of course if no problem is detected in a session, the segment corresponds to the whole session. We can also extract several segments in one session as is the case for  session 380 (Fig.~\ref{fig_session380}). The driving principle is to end up with segments as long as possible including an even number of orbital periods: $T=2n f_{\rm orb} $. Since $ f_{\rm spin}$ has been chosen such that $ f_{\rm spin}=(q/2) f_{\rm orb}$ ($q$ {odd} integer), all  potential signals at frequencies  $f_i=k_i f_{\rm orb} + p_i f_{\rm spin}$ (with $k_i $ and $p_i$ any integers) combining the orbital frequency $f_{\rm orb}$ and the spin frequency  $f_{\rm spin}$ are such that
\begin{equation}
  \label{eq:14}
 f_i= k_i f_{\rm orb} + p_i f_{\rm spin}=\left( k_i + \frac12 p_iq\right) f_{\rm orb}=\left( 2k_i +  p_iq\right) n \frac1T.
\end{equation}
This means that $f_i$ corresponds to a sampling frequency of the discrete Fourier transform and the correlation between two signals at frequencies $f_i$ and $f_j$ respecting the above property is null in theory and very low in practice.
Tables \ref{tab-segments-SUREF}  and \ref{tab-segments-SUEP} show the segments selected for our analysis. This comprises  13 segments totalling 598 orbits for  SUREF  and 19 segments totalling 1362 orbits for SUEP.

\begin{table}
\caption{\label{tab-segments-SUREF}Characteristics of the segments selected for our analysis of the SUREF data. The segment number corresponds to the session number extended by an index when there are more than one segment in the session. The duration is given as a multiple of orbital periods, remembering that this period is about 5946 s. The position of the segment in the session is indicated by the first and the last orbit which are  included in the segment. The fourth column indicates the percentage of data eliminated {from each segment} during the pre-processing (see Sect. \ref{sec:detect-elim-glitch}).} 

\begin{indented}
\lineup
\item[]\begin{tabular}{@{}*{3}{llll}}
\br                              
$\0\0$Segment &Duration &Position in the session &Percentage of data \cr 
$\0\0$number  &(orbits)    &  (orbits)                       &eliminated (glitches) \cr 
\mr
\0\0120-1&22 & 23 to 44    &4   \cr
\0\0120-2&64 & 57 to 120  &15 \cr
\0\0174    &86 & 34 to 119  &25 \cr
\0\0176    &62 & 1 to 62      &40 \cr
\0\0294    &76 & 18 to 93    &17 \cr
\0\0376-1&36 & 8 to 43      &14 \cr
\0\0376-2&28 & 52 to 79    &11 \cr
\0\0380-1&46 & 24 to 69    &7   \cr
\0\0380-2&34 & 75 to 108  &5   \cr
\0\0452    &32 & 1 to 32      &20 \cr
\0\0454    &56 & 1 to 56      &22 \cr
\0\0778-1&38 & 1 to 38      &0   \cr
\0\0778-2&18 & 41 to 58    &6  \cr 
\br
\end{tabular}
\end{indented}
\end{table}

\begin{table}
\caption{\label{tab-segments-SUEP}Same as Table \ref{tab-segments-SUEP} but for SUEP.} 

\begin{indented}
\lineup
\item[]\begin{tabular}{@{}*{3}{llll}}
\br                              
$\0\0$Segment &Duration &Position in the session &Percentage of data \cr 
$\0\0$number  &(orbits)    &  (orbits)                       &eliminated (glitches) \cr 
\mr
\0\0210    &50   & 1 to 50     &18 \cr
\0\0212    &60   & 1 to 60     &17 \cr
\0\0218    &120 & 1 to 120   &15 \cr
\0\0234    &92   & 1 to 92     &18 \cr
\0\0236    &120 & 1 to 120   &21 \cr
\0\0238    &120 & 1 to 120   &24 \cr
\0\0252    &106 & 1 to 106   &26 \cr
\0\0254    &120 & 1 to 120   &27 \cr
\0\0256    &120 & 1 to 120   &28 \cr
\0\0326-1&66   & 2 to 67     &12 \cr
\0\0326-2&34   & 69 to 102 &7  \cr
\0\0358    &92   & 1 to 92     &14 \cr
\0\0402    &18   & 3 to 20     &35 \cr
\0\0404    &120 & 1 to 120   &23 \cr
\0\0406    &20   & 1 to 20     &23 \cr
\0\0438    &32   & 1 to 32     &21 \cr
\0\0442    &40   & 1 to 40     &21 \cr
\0\0748    &24   & 1 to 24     &25 \cr
\0\0750    &8     & 1 to 8       &19 \cr
\br
\end{tabular}
\end{indented}
\end{table}

\subsection{Glitches}
\label{sec:singular-events}

\subsubsection{General characteristics of glitches.}
\label{sec:gener-char-glitch}

\begin{figure} 
\center
\includegraphics[width=0.7\textwidth]{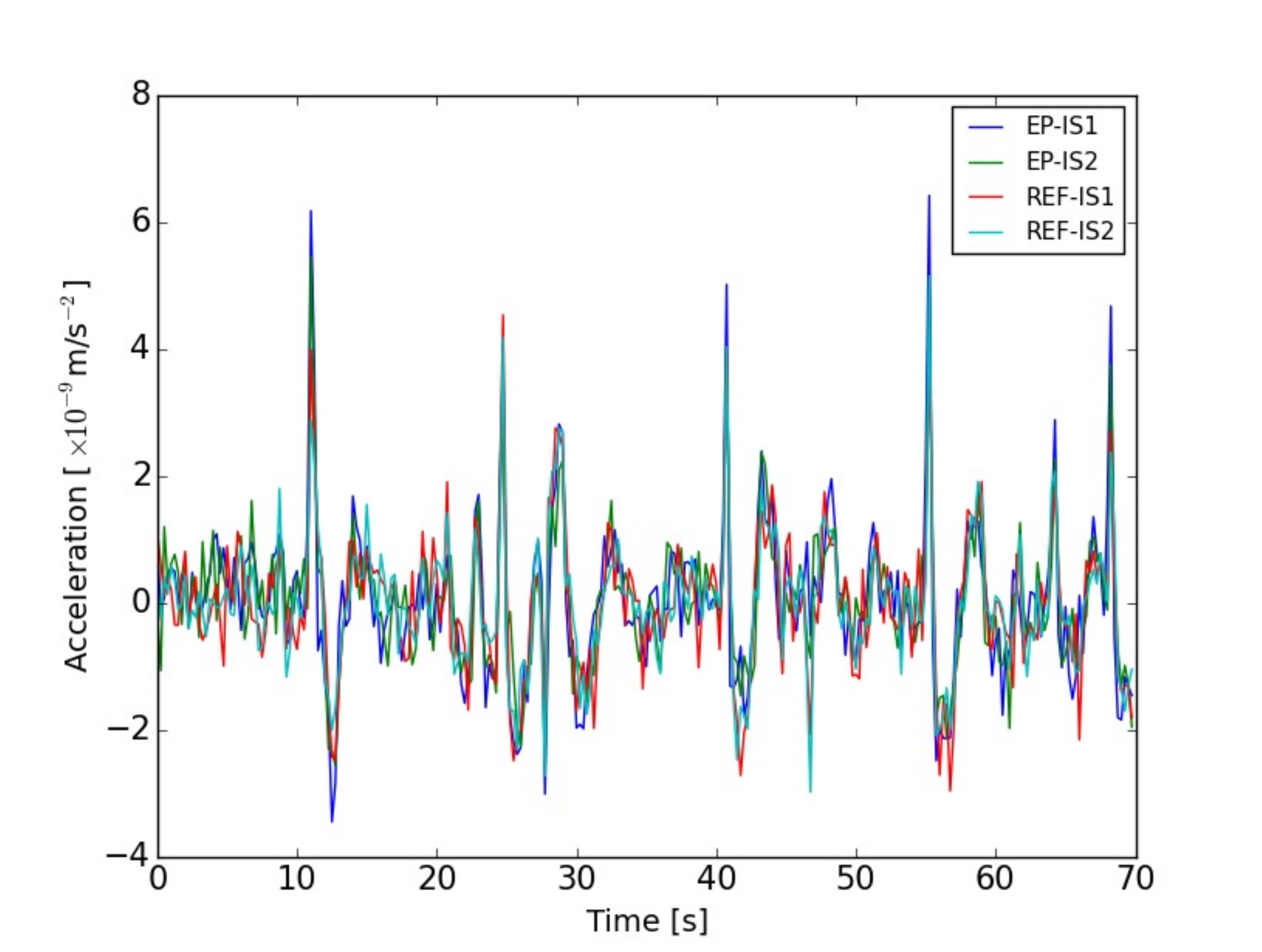}
\caption{Example of superposition of measured acceleration (after removing of a low frequency trend) by the 4 test-masses at the same time. The large peaks appear on all masses but with slightly different amplitudes.}
\label{fig_glitches4masses}       
\end{figure}

\begin{figure} 
\center
\includegraphics[width=0.7\textwidth]{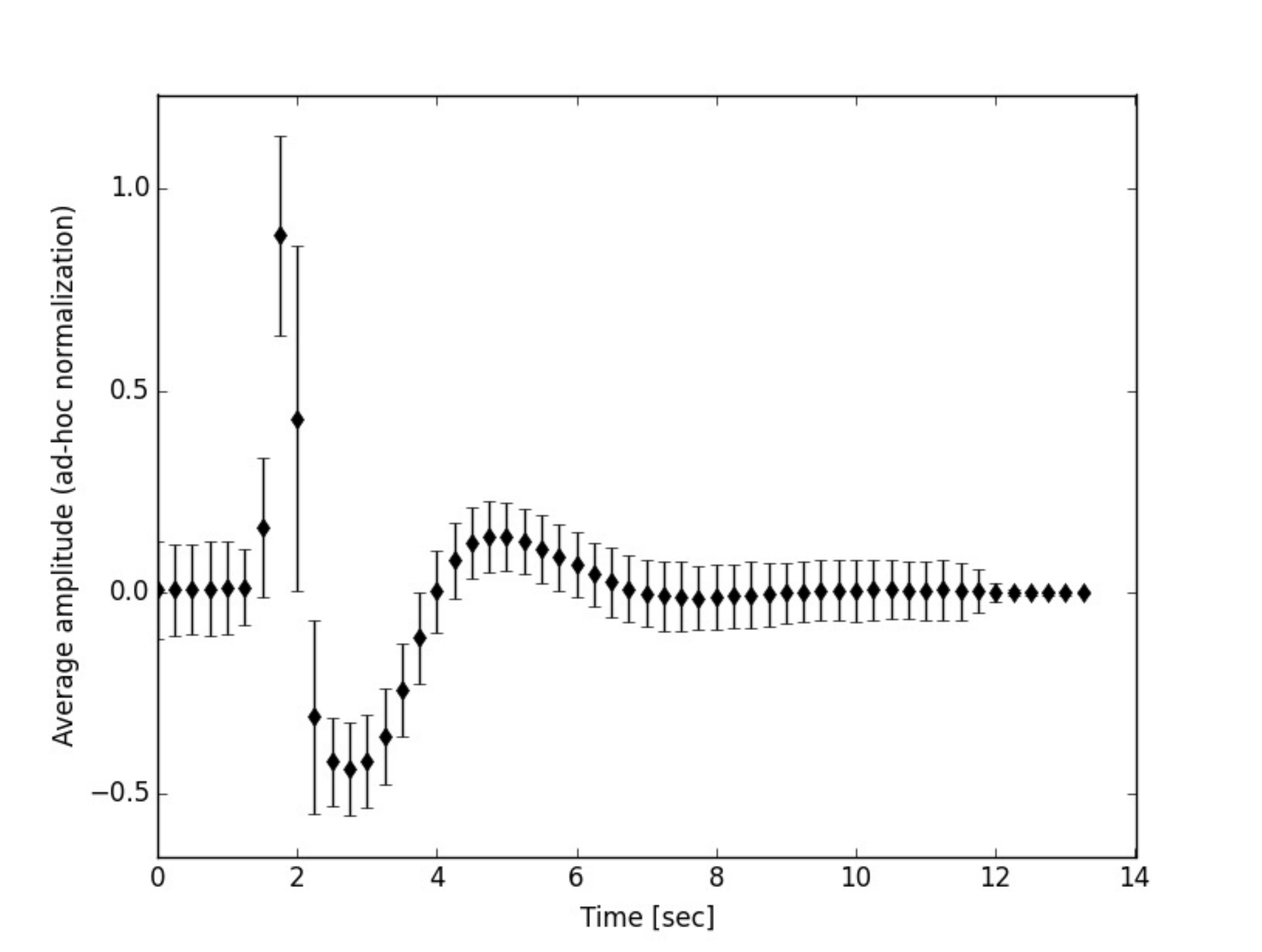}
\caption{Mean shape of observed glitches \cite{bergecqg8}}
\label{fig_GlitcheShape}       
\end{figure}

When looking closely at the temporal evolution of the measured accelerations on several test-masses, we can see short (a few seconds) and {significant} (1-10 nms$^{-2}$) variations which appear at the same time on both masses of the operating SU and even on the 4 masses when the 2 SUs are operating simultaneously (Fig.~\ref{fig_glitches4masses}). The simultaneous {appearance} of these features for  all masses proves that these events have the same common external source. This kind of event has been already observed in other space missions {carrying} accelerometers \cite{flury08}  and are called ``twangs'' or  ``glitches''.  Glitches in MICROSCOPE have been extensively studied in a dedicated paper \cite{bergecqg8}. Here, we recall their main characteristics: 
\begin{itemize}
\item As seen through the transfer functions of the drag-free and {of the} instrument, glitches look like exponentially damped sines with a large first ramp in one direction followed by a smoother oscillation in the opposite direction; their mean shape has been computed in  \cite{bergecqg8} and is shown in Fig.~\ref{fig_GlitcheShape}; even if the source events probably do not last more than a few milliseconds, they affect the measured acceleration for a dozen seconds.
\item The amplitude of the corresponding measured acceleration can reach up to a few $10^{-8}\,$ms$^{-2}$ but can be much smaller; there are probably also glitches which are masked by the measurement noise.
\item The number of detected glitches {typically} ranges from 0.02 to 0.06 s$^{-1}$.
\item Although they can occur at any time, their probability of occurrence is affected by two periodicities: the orbital period of the satellite and its spin period.
\end{itemize}
Their most likely origins  are {crackling} of the MLI (Multi-Layer Insulation) of the satellite and more rarely clangs of the gas tanks used for the micro-propulsion. Predicting the exact occurrence, form and amplitudes of the glitches seems out of reach. Moreover, although very similar, the responses of the different test masses to these sudden events are not perfectly identical and the corresponding signal is not fully cancelled in the differential acceleration \cite{bergecqg8}. Thus, due to the time distribution of the glitches, they could generate a tiny signal at  the $f_{\rm EP}$ frequency. Consequently, the chosen strategy is to detect and eliminate them.

\subsubsection{Detection and elimination of the glitches.}
\label{sec:detect-elim-glitch}

Glitches are detected in a {double} two steps procedure \cite{rodriguescqg4, bergecqg7, bergecqg8}: (i) we use a standard recursive $\sigma$-clipping technique (e.g. Ref.~\cite{akhlaghi15}) to extract outliers ($4.5\sigma$) from the measured differential accelerations on the three axes ($X,Y,Z$) simultaneously before (ii) flagging data points in the second preceding the outliers and the 15 seconds following it  (typically, a single glitch builds from the noise within 0.5 seconds and dies off within 5 to 10 seconds from its peak, depending on its signal-to-noise ratio). Thence  we  define a first mask $M_1$ made of zeros in segments characterised as glitches and ones elsewhere.
The same two steps are then performed on the high-frequency-filtered differential acceleration (using a 2$^{\rm nd}$ order Butterworth filter of critical frequency 0.01~Hz), which allows us to detect low signal-to-noise glitches and define a second mask $M_2$ (a $3\sigma$ threshold is used during these steps). The final mask is the logical sum of both masks, $M=M_1\times{}M_2$. The percentage of masked data for each session is indicated in Tables \ref{tab-segments-SUREF} and \ref{tab-segments-SUEP}. These are typically 20\% but with less than 10\% for some sessions and 40\% for session 176.

\section{Separated analysis of each segment}
\label{sec:details-analysis}
As {explained} in Section \ref{sec:segmentation} we have 13 segments for SUREF and 19 segments for SUEP. A first important objective is to analyse these segments separately since, as discussed in Refs.~\cite{touboul17,touboul19} the analysis of a single segment already {resulted in an accuracy of estimation} of the E\"otv\"os parameter 10 times better than the previous experiments. Moreover this analysis will provide some insights into the data before a global analysis of all segments.

\subsection{Main steps of the analysis}
\label{sec:main-steps-analysis}

\begin{figure}
\centering
\includegraphics[width=0.80\textwidth]{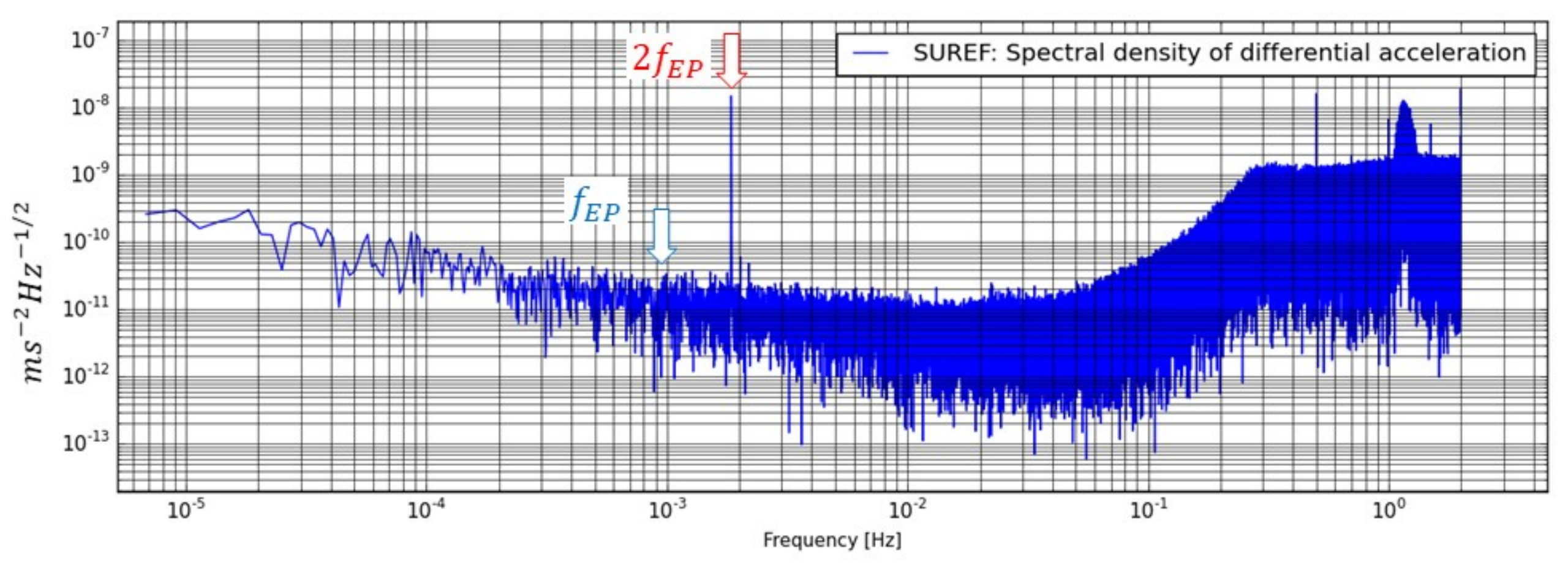}
\includegraphics[width=0.80\textwidth]{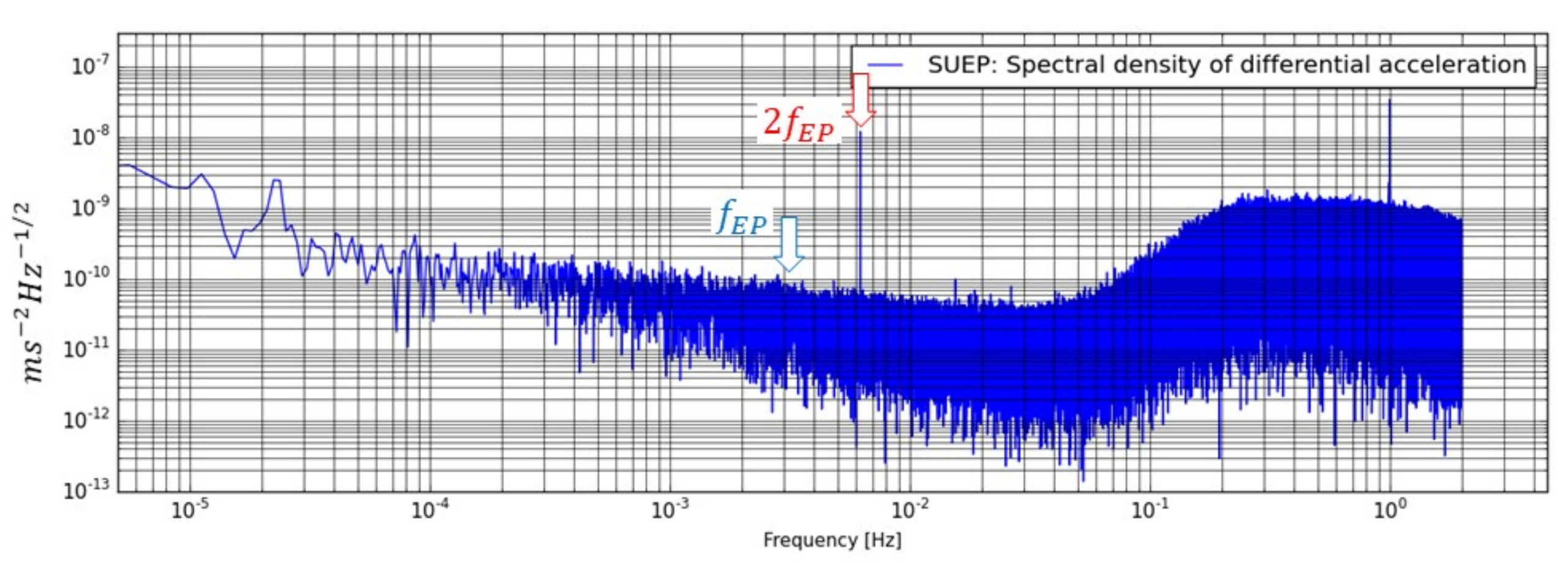}
\caption{{Amplitude} Spectral Density of the sensor differential acceleration along the $X$-axis for SUREF during session 176 (upper panel) and for SUEP during session 236 (lower panel). Note the {peak} at frequency $2 f_{\rm EP}$ coming from the Earth's gravity gradient due to the offcentring of the two test masses.}
\label{fig_noise}       
\end{figure}

{The analysis of individual segments is performed in the following steps:}
\begin{enumerate}
\item Calibration correction: {thanks to dedicated calibration sessions, it was possible to estimate the values of the instrumental parameters ${\Delta'_{y}}$, $a_{d11}$, $a_{d12}$ and $a_{d13}$ throughout the mission}~\cite{hardycqg6}; {using the values of  $S_{ij}$ and $\dot\Omega_i $  precisely computed for each measurement date and of the common mode measured acceleration $\tilde\Gamma^{(c)}_i$,  $\Gamma_x^{(d)}$ is corrected from the terms  ${\Delta'_{y}}\left({{ S_{xy}}}+\dot\Omega_z \right)$ and $ 2\left( { a_{d11}}{ \tilde\Gamma^{(c)}_x}+{ a_{d12}}{ \tilde\Gamma^{(c)}_y}+{ a_{d13}}{ \tilde\Gamma^{(c)}_z}\right)$}~\cite{bergecqg7}.
\item Detection of the  glitches: we detect glitches  and define a corresponding mask according to the algorithm described in section \ref{sec:detect-elim-glitch}. The union of this mask with the mask generated by the few points (typically a dozen per session) tagged directly on board {is} used during the analysis (see Sect.~\ref{sec:estim-param-time}) to discard the corresponding points. {Note that this detection is performed only for EP sessions since we have checked that this operation has no impact on the parameters estimated using the calibration sessions.}
  \item {The parameters $\alpha_j$, $\delta_x$, $\delta_z$, $\Delta'_{x}$ and $\Delta'_{z}$ are estimated by fitting the corrected measured differential acceleration to the model}\eqref{eq:2}.
\end{enumerate}
Since the measurement noise in MICROSCOPE is coloured (Fig.~\ref{fig_noise} and \cite{touboul19}), the optimal estimation of the parameters requires use of the characteristics of the noise. However, masking introduces gaps into the data, which are not regularly sampled any more. In this case, straightforward techniques {(periodograms)} to compute the Power Spectral Density (PSD) are  no longer effective: a {significant} leakage {of the signal power from high to low frequency bands,} leads to a substantial overestimation of the standard deviation of the estimated parameters \cite{baghi15, baghi16}. {To solve this problem,} we performed our estimation with two very different techniques described below: a method of maximisation of the likelihood in the time domain called M-ECM { (Modified Expectation Conditional Maximisation)} and a weighted least-squares regression in the Fourier domain named ADAM {(Accelerometric Data Analysis for MICROSCOPE)}. 

\subsection{Estimation of the parameters in the time domain using the M-ECM analysis}
\label{sec:estim-param-time}

M-ECM \cite{baghi16} is an inference algorithm designed to perform linear regression {of} gapped data. {Although M-ECM may find applications in many areas, it was developed by the MICROSCOPE team during the mission design for the purpose of processing the mission data.} Assuming a linear model for the signal and a stationary Gaussian distribution for the noise, it maximises the likelihood by iterating {the following} two steps. The first one is the conditional expectation step, which computes the expected likelihood (or rather, its logarithm $l_Y(\boldsymbol{\theta})$) conditionally to the observed data. This process amounts to estimating missing data within gaps. The second one is the maximisation step, which maximises the expected likelihood over the regression parameters $\boldsymbol{\theta}$. This step amounts to computing the generalised least-squares estimate of the parameters. It also includes estimating the noise PSD (the inclusion of this step {motivates the qualification} ``modified'' to the algorithm's name, as it is not present in standard ECM algorithms~\cite{Dempster1977, MengRubin1993}). The use of the M-ECM algorithm allows {one to avoid} spectral leakage effects due to data gaps by restoring the statistics of the noise in the frequency domain.

In the case of complete data, and under the stationary assumption, we can write the logarithm of the likelihood as
\begin{equation}
	l_Y(\boldsymbol{\theta}) \approx - \frac{1}{2} \left\{\log \left(\mathrm{det}\left[\boldsymbol{\gamma}\right]\right)  + \left( \hat{\mathbf{Y}} - \left[\hat{\mathbf{A}}\right]\boldsymbol{\theta}  \right)^{\dagger} \left[\boldsymbol{\gamma}\right]^{-1} \left(\hat{\mathbf{Y}} - \left[\hat{\mathbf{A}}\right]\boldsymbol{\theta}\right) \right\},
	\label{eq:log-likelihood-fourier}
\end{equation} 
 where ${\hat{\mathbf Y}} $ is the vector of $N$ Fourier-transformed measurements, ${\boldsymbol{\theta}}$ is the vector of parameters to estimate and $\left[\hat{\mathbf{A}}\right]$ is the matrix of derivatives of the model with respect to these parameters. The matrix $\left[\boldsymbol{\gamma}\right]$ denotes the covariance of the noise in the Fourier domain. It is approximately diagonal {(see e.g. Fig.~1 of Ref.}~\cite{ bergecqg7}) and its diagonal elements are equal to the noise PSD. If all measurements are available, we solve the estimation problem by maximising the likelihood with respect to $\boldsymbol{\theta}$. 
 
When data points are missing, directly maximising Eq.~(\ref{eq:log-likelihood-fourier}) can be computationally cumbersome because we cannot approximate  $\left[\boldsymbol{\gamma}\right]$ as a diagonal matrix. Instead, the estimation is broken down in two steps. In the first step, we compute the expectation of the log-likelihood given the observed data $\mathbf{Y}_{o}$ and the value of the parameters at the present iteration:
\begin{equation}
	\text{E step:   }\, Q_Y(\boldsymbol{\theta}, \boldsymbol{\gamma} ) = \operatorname{E}\left[ l_Y(\boldsymbol{\theta}) | \mathbf{Y}_{o}, \boldsymbol{\theta}, \boldsymbol{\gamma} \right].
\end{equation}
This computation is called the expectation step (E). It requires {computing} the expectation of the full data vector conditionally on the observed data $\operatorname{E}\big[ \hat{\mathbf{Y}} | \mathbf{Y}_{o}, \boldsymbol{\theta},  \boldsymbol{\gamma} \big]$, along with its second-order moment $\operatorname{E}\big[ \hat{\mathbf{Y}} \hat{\mathbf{Y}}^{\dagger}  | \mathbf{Y}_{o}, \boldsymbol{\theta},  \boldsymbol{\gamma} \big]$. We can compute these quantities by using the conditional mean and covariance formulas for Gaussian processes (see~\cite{baghi16} for more details). 

The second step is similar to the maximisation one would do for complete data, except that now we maximise the expected likelihood:
\begin{equation}
	\text{M step:   }\, \boldsymbol{\theta} = \operatorname*{argmax}_\theta Q_Y\left(\boldsymbol{\theta}, \boldsymbol{\gamma} \right).
\end{equation}
This step is the maximisation (M) step. Note that solving for $\boldsymbol{\theta}$ is done for {a given} noise PSD $\boldsymbol{\gamma}$. In the M-ECM algorithm, we assume that $\boldsymbol{\gamma}$ is unknown and that it depends on some noise parameters $\boldsymbol{\beta}$. Hence, we also fit for the PSD by performing a pseudo maximisation conditionally on $\boldsymbol{\theta}$, so that
\begin{equation}
   \boldsymbol{\beta} = \operatorname*{argmax}_\beta Q_Y\left(\boldsymbol{\theta}, \boldsymbol{\gamma}(\boldsymbol{\beta}) \right).
\end{equation}

Finally, we iterate E and M steps until $\boldsymbol{\theta}$ converges. The final solution is the value that maximises the likelihood with respect to observed data $l_{Y_{o}}(\boldsymbol{\theta})$. {As a result, M-ECM first computes the expectation of the likelihood through data reconstruction and then maximises it over the parameters. Hence, the maximisation takes advantage of the fast Fourier transform applied to the regularly sampled reconstructed time series, which is computationally more efficient than the direct maximisation of the gapped data likelihood.} Values drawn from the missing data conditional distribution (also called reconstructed data in the following) is a useful by-product of M-ECM.

{The resulting algorithm is unbiased after several iterations because it converges to the same solution as the one obtained with a direct (but costly) maximisation of the gapped, time-domain data likelihood. M-ECM is also approximately optimal in the statistical sense, as it yields the solution with nearly minimal variance, provided that the PSD estimation is sufficiently accurate\cite{baghi16}.}

\subsection{Estimation of the parameters in the Fourier domain using the ADAM software}
\label{sec:estim-param-four}

{ADAM was developed by the MICROSCOPE team during the mission preparation for the purpose of processing the data of the experiment.}
{For the estimation with ADAM}, we use both original data remaining after the masking operation and data reconstructed by M-ECM as described in the previous section. This means that we have now data without gaps, i.e. regularly sampled. This allows us to apply a discrete Fourier transform (DFT) converting the whole system of measurement equation in the frequency domain: whereas in the time domain each equation is associated to a time $t_i$, each equation of the transformed system is associated to a frequency $f_j$. This leads to several interesting properties and in particular:
\begin{itemize}
\item In case of periodic signals, their energy {is} concentrated in a small number of {frequencies, corresponding to a small number of} equations in the frequency domain; this is the case of the gravity acceleration and of the gravity gradient; moreover thanks to our choice to impose $f_{\rm spin}=k f_{\rm orb}/2$ and to perform the analysis over $2n T_{\rm orb}$ {($k$ and $n$ integers)}, the frequencies $ f_{\rm orb}$,  $f_{\rm spin}$ and  $f_{\rm EP}$ correspond precisely to a frequency of the DFT:  $f_q=q/T_{\rm analysis}=q  f_{\rm orb}/(2n)$ with $q_{\rm orb}=2n$, $q_{\rm spin}=kn$ and $q_{\rm EP}=(k+2)n$. 
\item {With long enough data streams, the covariance is approximately diagonal, with a squared error inversely proportional to the data size}~\cite{Zhu2017}. {In our application, the deviation is below the percent level.} Thus a diagonal weighting matrix composed of the elements $ w(f_k)=\frac{1}{\sqrt{\gamma(f_k)}}$ where $\gamma(f_k)$ is the PSD of the noise at the
frequency $f_k$~\cite{bergecqg7}, is almost optimal. 
  \end{itemize}
{Details of the procedure} {are} described in \cite{bergecqg7} and we recall here the main steps:
\begin{enumerate}
 \item The series $\Gamma^{(d)}_{x, {\rm corr}}(t_i)$, $(t_i-t_0)^j$, $g_i(t_i)$, $g_z(t_i)$, $S_{xx} (t_i)$ and $S_{xz} (t_i)$ are transformed in the frequency domain by application of a DFT: the $N$ observation equations  in the time domain {(corresponding to} Eq.~\eqref{eq:2} {at $N$ different times)} are transformed into $N$ observation equations in the frequency domain.
 \item Potentially, we select only a subset of equations in the frequency domain, which is equivalent to selecting frequency bands. For the standard analysis, the frequency band around $f_{\rm EP}$ is selected to estimate $\delta_x$ and $\delta_z$, and the frequency band around $2 f_{\rm EP}$ is selected to estimate $\Delta'_{x}$ and $\Delta'_{z}$. We choose a bandwidth large enough to encompass all the relevant signals (gravity acceleration and gravity gradient with their significant harmonics as well as the rotational terms): $8\times{}10^{-4}$ Hz for sessions in spin V2 and $2\times{}10^{-3}$ Hz for sessions in spin V3. The whole spectrum is used only to estimate the parameters of the low frequency trend described above. 
 \item These observation equations are used to estimate the parameters of Eq.~\eqref{eq:2} and the associated statistical errors using a weighted least-squares inversion.
 \end{enumerate}
Fundamentally, no new information is expected from the ADAM analysis that performs the parameters estimation in the Fourier domain using the reconstructed data provided by M-ECM. However, it is interesting to cross-check results from different methods. 
Moreover, ADAM is much faster than M-ECM: for a segment of 120 orbits M-ECM needs about 12 hours while ADAM takes only a few minutes. This is because in the E-step, M-ECM has to solve a large linear system where the system matrix is the covariance of the observed data in the time domain. The preconditioning is also memory-expensive. {Although these disadvantages are not critical} for a single run, they become a serious problem for the numerous tests required to strengthen the quality and the robustness of the estimation analysis.

\subsection{Results}
\label{sec:results}

\begin{figure}
\centering
\includegraphics[width=0.80\textwidth]{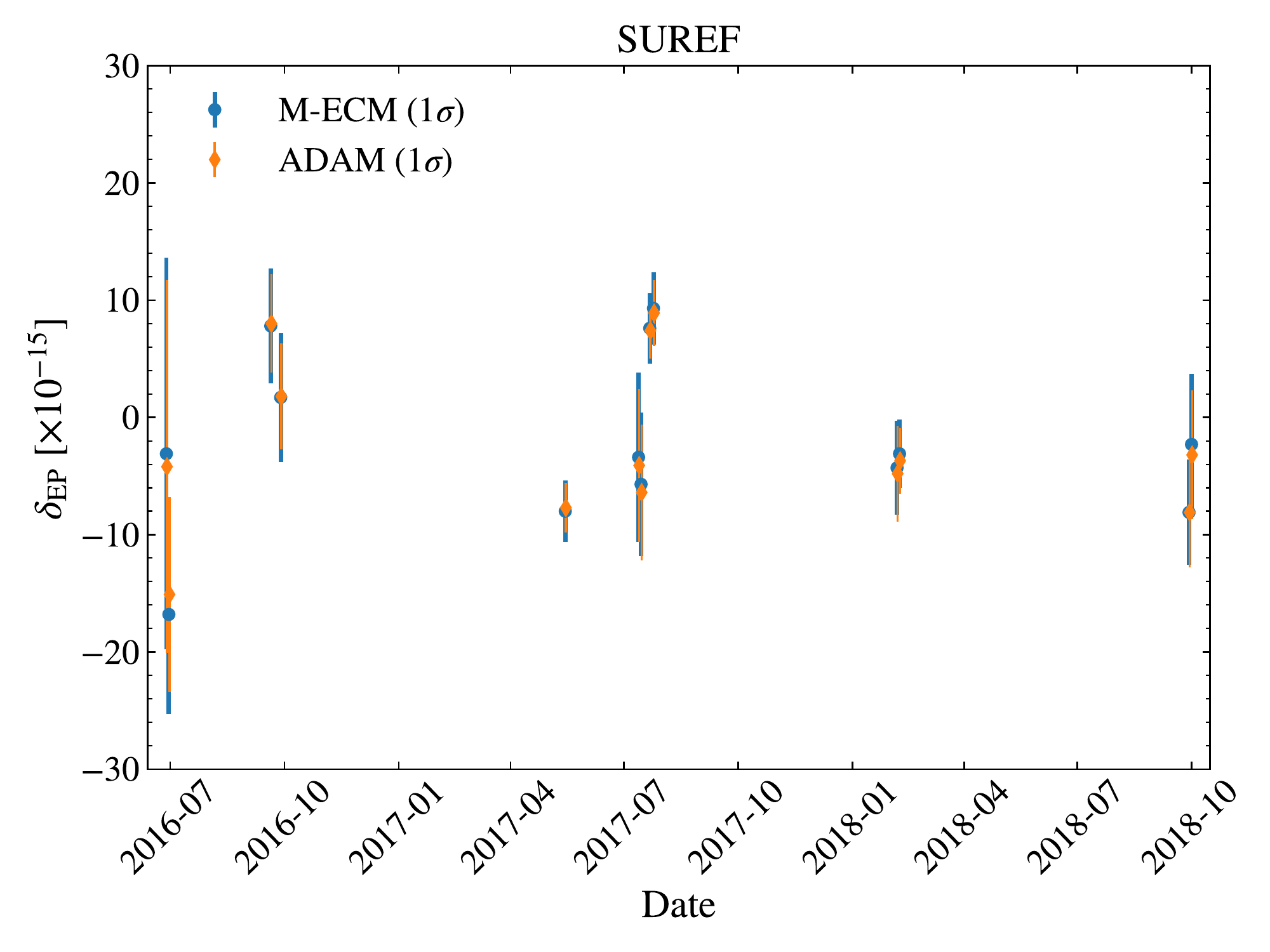}
\caption{E\"otv\"os parameter estimates for each SUREF segment and their 68\% confidence error bars. Blue circles show M-ECM's estimates and orange ones {\sc Adam}'s.}
\label{fig_delta_SUREF}       
\end{figure}

\begin{figure}
\centering
\includegraphics[width=0.80\textwidth]{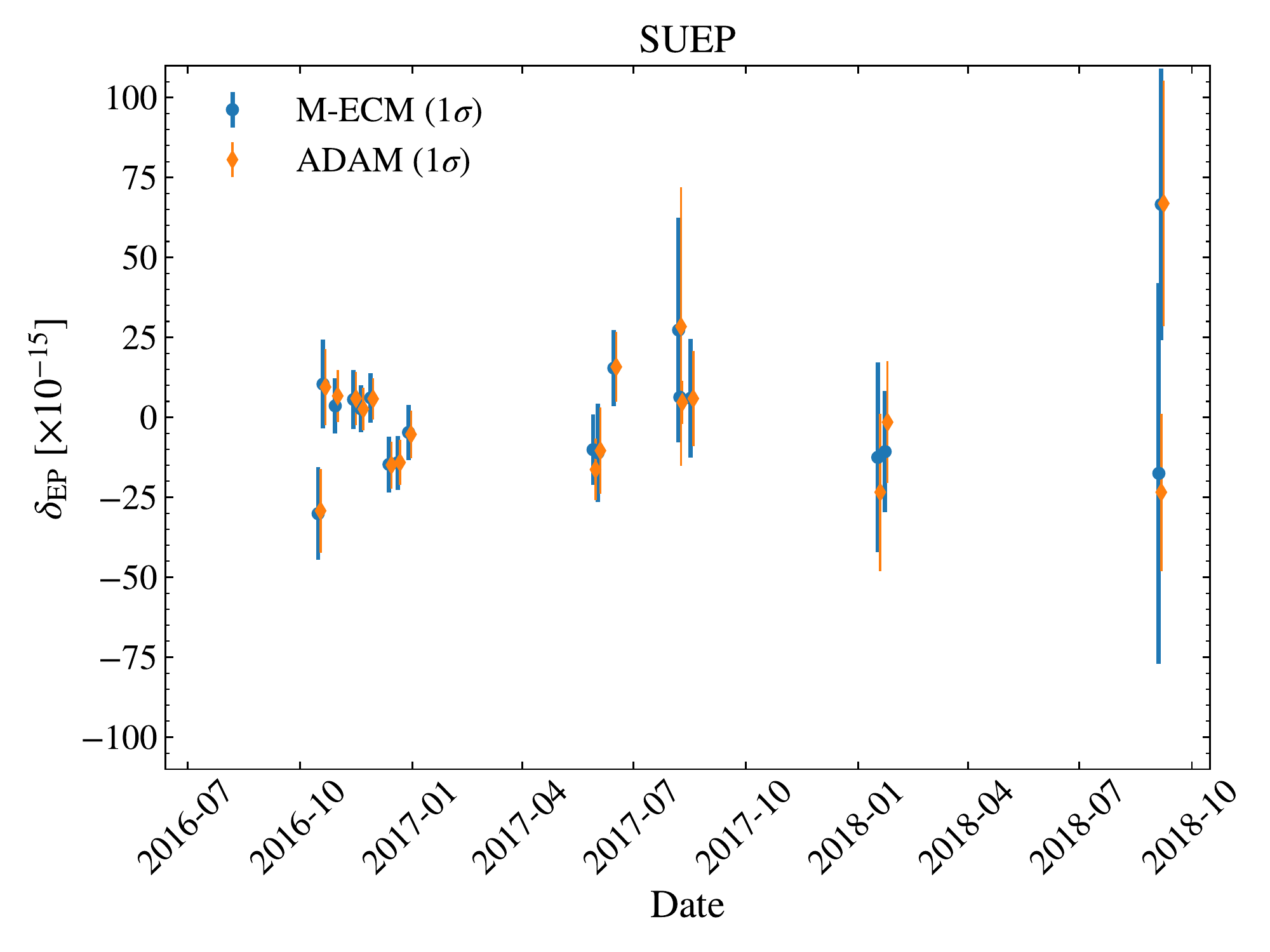}
\caption{Same as Fig. \ref{fig_delta_SUREF} but for SUEP.}
\label{fig_delta_SUEP}       
\end{figure}

The {estimates of the parameters $\delta_x$ and of  $\delta_z$ and their standard deviations are listed} in Tables \ref{tab-estimations-SUREF} and \ref{tab-estimations-SUEP}.
{As noted in} Sect.~\ref{sec:measurement-model}, {$\delta_z =\tilde{a}_{c13} \delta$  is almost three orders of magnitude smaller than $\delta$, i.e. far below what is detectable; thus the estimated values of $\delta_z$ and of its standard deviation are indicators of statistical or systematic effects.} Figs.~\ref{fig_delta_SUREF} and \ref{fig_delta_SUEP} {give an overview of the estimates of the E\"otv\"os parameter with their time distribution.}
The values of the estimated offcentring are reported and commented in \cite{hardycqg6}.

\begin{table}
  \caption{\label{tab-estimations-SUREF}Values of the E\"otv\"os parameter $\delta_x$with their associated standard deviation estimated from SUREF measurements over {individual}  segments. The table reports also the value of the component  $\delta_z$ in phase quadrature. The values followed by (M) result from the M-ECM analysis while the values followed by (A) come from ADAM. Segments marked by an asterisk $^*$ correspond to sessions in spin V3 and the others to sessions in spin V2.} 

\lineup
\begin{tabular}{@{}lrrrrrr}
\br                              
$\0\0$Segment & $\delta_x$ (M)& $\delta_x$ (A) & $\delta_z$ (M)& $\delta_z$ (A) &$\delta_x/\sigma$ (M)&$\delta_x/\sigma$ (A)\cr 
$\0\0$number&$\times{}10^{15} $ &$\times{}10^{15} $ &$\times{}10^{15} $ &$\times{}10^{15} $  &           &            \cr 
\mr
\0\0{120-1}        &{-3.1}\ $\pm{}$\,{16.7}                              &{-4.2}\ $\pm{}$\,{15.9}                           &{13.5}\ $\pm{}$\,{16.7}                             &{10.3}\ $\pm{}$\,{15.9}                            & {-0.2}  & {-0.3}    \cr    
\0\0{120-2}        &{-16.8}\ $\pm{}$\,\phantom{5}{8.5}          &{-15.1}\ $\pm{}$\,\phantom{5}{8.3}       &{-7.6}\ $\pm{}$\,\phantom{5}{8.5}           &{-7.5}\ $\pm{}$\,\phantom{5}{8.3}           & {-2.0}  & {-1.8}    \cr    
\0\0{174}            &{7.8}\ $\pm{}$\,\phantom{5}{4.9}             &{8.0}\ $\pm{}$\,\phantom{5}{4.2}           &{-13.5}\ $\pm{}$\,\phantom{5}{4.9}        &{-14.0}\ $\pm{}$\,\phantom{5}{4.2}           & {1.6}    & {1.9}      \cr
\0\0{176}            &{1.7}\ $\pm{}$\,\phantom{5}{5.5}             &{1.8}\ $\pm{}$\,\phantom{5}{4.5}            &{7.9}\ $\pm{}$\,\phantom{5}{5.5}            &{8.4}\ $\pm{}$\,\phantom{5}{4.5}              & {0.3}    & {0.4}      \cr
\0\0{294$^*$}     &{-8.0}\ $\pm{}$\,\phantom{5}{2.6}           & {-7.7}\ $\pm{}$\,\phantom{5}{2.1}         &  {-2.8}\ $\pm{}$\,\phantom{5}{2.6}         &{-2.3}\ $\pm{}$\,\phantom{5}{2.1}         & {-3.1}  & {-3.6}    \cr   
\0\0{376-1}        &{-3.4}\ $\pm{}$\,\phantom{5}{7.2}            &{-4.1}\ $\pm{}$\,\phantom{5}{6.5}         &{-8.5}\ $\pm{}$\,\phantom{5}{7.2}           &{-7.9}\ $\pm{}$\,\phantom{5}{6.5}          & {-0.5}  & {-0.6}    \cr  
\0\0{376-2}        &{-5.7}\ $\pm{}$\,\phantom{5}{6.1}           &{-6.4}\ $\pm{}$\,\phantom{5}{5.8}         &{14.5}\ $\pm{}$\,\phantom{5}{6.1}          &{16.1}\ $\pm{}$\,\phantom{5}{5.8}            & {-0.9}  & {-1.1}    \cr  
\0\0{380-1$^*$} & {7.6}\ $\pm{}$\,\phantom{5}{3.0}            & {7.4}\ $\pm{}$\,\phantom{5}{2.4}          &{-10.1}\ $\pm{}$\,\phantom{5}{3.0}        &{-10.2}\ $\pm{}$\,\phantom{5}{2.4}          & {2.5}    & {3.1}      \cr   
\0\0{380-2$^*$} &{9.3}\ $\pm{}$\,\phantom{5}{3.1}              &{8.9}\ $\pm{}$\,\phantom{5}{2.8}           &{-9.1}\ $\pm{}$\,\phantom{5}{3.1}           &{-9.2}\ $\pm{}$\,\phantom{5}{2.8}           & {3.0}    & {3.2}      \cr 
\0\0{452}            &{-4.3}\ $\pm{}$\,\phantom{5}{4.0}           &{-4.8}\ $\pm{}$\,\phantom{5}{4.1}          &{14.6}\ $\pm{}$\,\phantom{5}{4.0}           &{16.1}\ $\pm{}$\,\phantom{5}{4.1}           & {-1.1}  & {-1.2}    \cr
\0\0{454}            &{-3.1}\ $\pm{}$\,\phantom{5}{2.9}           &{-3.7}\ $\pm{}$\,\phantom{5}{2.8}          &{19.9}\ $\pm{}$\,\phantom{5}{2.9}          &{19.8}\ $\pm{}$\,\phantom{5}{2.8}            & {-1.1}  & {-1.3}     \cr
\0\0{778-1}        &{-8.1}\ $\pm{}$\,\phantom{5}{4.5}            &{-8.1}\ $\pm{}$\,\phantom{5}{4.7}          &{22.8}\ $\pm{}$\,\phantom{5}{4.5}           &{22.6}\ $\pm{}$\,\phantom{5}{4.7}         & {-1.8}  & {-1.7}     \cr 
\0\0{778-2}        &{-2.3}\ $\pm{}$\,\phantom{5}{6.0}            &{-3.2}\ $\pm{}$\,\phantom{5}{5.5}          &{20.0}\ $\pm{}$\,\phantom{5}{6.0}           &{18.6}\ $\pm{}$\,\phantom{5}{5.5}         & {-0.4}  & {-0.6}    \cr  
\br
\end{tabular}
\end{table}

\begin{table}
\caption{\label{tab-estimations-SUEP}Same as Table \ref{tab-estimations-SUREF} but for SUEP.} 

\lineup
\begin{tabular}{@{}lrrrrrr}
\br                              
Segment & $\delta_x$ (M)& $\delta_x$ (A) & $\delta_z$ (M)& $\delta_z$ (A) &$\delta_x/\sigma$ (M)&$\delta_x/\sigma$ (A)\cr 
number&$\times{}10^{15} $ &$\times{}10^{15} $ &$\times{}10^{15} $ &$\times{}10^{15} $  &           &            \cr 
\mr
{210$^*$}    & {-30.1}\ $\pm{}$\,{14.5}                          &{-29.2}\ $\pm{}$\,{13.1}                    & {-25.1}\ $\pm{}$\,{14.5}                      & {-25.3}\ $\pm{}$\,{13.1}                     &-2.1  & {-2.2}  \cr
{212$^*$}    &{10.4}\ $\pm{}$\,{13.9}                             &{9.5}\ $\pm{}$\,{11.9}                        & {1.0}\ $\pm{}$\,{13.9}                          & {1.3}\ $\pm{}$\,{11.9}                         &0.7     & {0.8}  \cr
{218$^*$}    & {3.6}\ $\pm{}$\,\phantom{5}{8.7}            & {6.7}\ $\pm{}$\,\phantom{5}{8.1}     & {8.2}\ $\pm{}$\,\phantom{5}{8.7}         &{7.5}\ $\pm{}$\,\phantom{5}{8.1}        & 0.4   & {0.8}  \cr
{234$^*$}    & {5.6}\ $\pm{}$\,\phantom{5}{9.3}           & {5.9}\ $\pm{}$\,\phantom{5}{8.3}     & {-4.2}\ $\pm{}$\,\phantom{5}{9.3}       & {-4.6}\ $\pm{}$\,\phantom{5}{8.3}      &0.6    & {0.7}  \cr
{236$^*$}    & {2.7}\ $\pm{}$\,\phantom{5}{7.4}           & {2.6}\ $\pm{}$\,\phantom{5}{6.6}     & {16.7}\ $\pm{}$\,\phantom{5}{7.4}        &{16.8}\ $\pm{}$\,\phantom{5}{6.6}      &  0.4   & {0.4}   \cr
{238$^*$}    & {6.1}\ $\pm{}$\,\phantom{5}{7.8}           & {5.8}\ $\pm{}$\,\phantom{5}{6.4}     &{-3.9}\ $\pm{}$\,\phantom{5}{7.8}        & {-3.7}\ $\pm{}$\,\phantom{5}{6.4}      & 0.8    & {0.9}  \cr
{252$^*$}    & {-14.7}\ $\pm{}$\,\phantom{5}{8.7}       &{-14.9}\ $\pm{}$\,\phantom{5}{7.3}   &{1.8}\ $\pm{}$\,\phantom{5}{8.7}          & {2.4}\ $\pm{}$\,\phantom{5}{7.3}       & -1.7  & {-2.0} \cr
{254$^*$}    & {-14.2}\ $\pm{}$\,\phantom{5}{8.4}      & {-14.1}\ $\pm{}$\,\phantom{5}{7.0}  & {-24.9}\ $\pm{}$\,\phantom{5}{8.4}      & {-25.8}\ $\pm{}$\,\phantom{5}{7.0}   & -1.7   & {-2.0} \cr
{256$^*$}    &{-4.7}\ $\pm{}$\,\phantom{5}{8.6}         & {-5.3}\ $\pm{}$\,\phantom{5}{7.4}    & {14.1}\ $\pm{}$\,\phantom{5}{8.6}        & {13.3}\ $\pm{}$\,\phantom{5}{7.4}     & -0.5  & {-0.7}  \cr
{326-1$^*$}& {-10.1}\ $\pm{}$\,{11.1}                        &{-16.3}\ $\pm{}$\,\phantom{5}{9.6}   &{3.7}\ $\pm{}$\,{11.1}                           &{2.3}\ $\pm{}$\,\phantom{5}{9.6}          &-0.9   & {-1.7}  \cr
{326-2$^*$}&{-11.1}\ $\pm{}$\,{15.4}                        & {-10.4}\ $\pm{}$\,{13.5}                     & {5.5}\ $\pm{}$\,{15.4}                          &{5.3}\ $\pm{}$\,{13.5}                          & -0.7  &  {-0.8} \cr
{358$^*$}    &{15.4}\ $\pm{}$\,{11.9}                          &{15.8}\ $\pm{}$\,{10.9}                       &{-2.0}\ $\pm{}$\,{11.9}                          & {-1.9}\ $\pm{}$\,{10.9}                       &1.3    & {1.4}  \cr
{402}           & {27.3}\ $\pm{}$\,{35.1}                         &{ 28.4}\ $\pm{}$\,{43.6}                    & {19.1}\ $\pm{}$\,{35.1}                         & {28.7}\ $\pm{}$\,{43.6}                      &0.8    & {0.7}  \cr
{404$^*$}    & {6.3}\ $\pm{}$\,{\phantom{5}7.9}          & {4.7}\ $\pm{}$\,\phantom{5}{6.7}     & {-5.9}\ $\pm{}$\,\phantom{5}{7.9}         & {-5.0}\ $\pm{}$\,\phantom{5}{6.7}    & 0.8   &0.7    \cr
{406$^*$}    & {6.0}\ $\pm{}$\,{18.6}                           & {5.9}\ $\pm{}$\,{14.9}                       &{44.0}\ $\pm{}$\,{18.6}                           &{44.1}\ $\pm{}$\,{14.9}                        &0.3    & {0.4} \cr
{438}           &{-12.5}\ $\pm{}$\,{29.6}                       & {-23.4}\ $\pm{}$\,{24.6}                  & {62.9}\ $\pm{}$\,{29.6}                         &{54.5}\ $\pm{}$\,{24.6}                        &-0.4  & -0.9    \cr
{442}           &{-10.7}\ $\pm{}$\,{19.0}                       & {-1.5}\ $\pm{}$\,{19.1}                    & {-5.2}\ $\pm{}$\,{19.0}                        &{-6.3}\ $\pm{}$\,{19.1}                         & -0.6 &-0.1     \cr
{748}           &{-17.5}\ $\pm{}$\,{59.6}                       & {-23.4}\ $\pm{}$\,{24.6}                  & {0.5}\ $\pm{}$\,{59.6}                          &{54.5}\ $\pm{}$\,{24.6}                         &-0.3  &-0.9    \cr
{750$^*$}    & {66.6}\ $\pm{}$\,{42.4}                         & {66.9}\ $\pm{}$\,{38.4}                     & {23.6}\ $\pm{}$\,{42.4}                         &{23.5}\ $\pm{}$\,{38.4}                          &1.6     & {1.7} \cr
\br
\end{tabular}
\end{table}

Several comments are in order.
\begin{enumerate}
\item Although being very different algorithms, M-ECM and ADAM provide very similar results both for the values of $\delta_x$ and $\sigma$. The only {appreciable} difference is for $\delta_z$ estimated for  segment 748 which is a SUEP session  in spin V2 lasting 24 orbits; but this difference is statistically {insignificant} since it is smaller than the standard deviation computed by M-ECM.
\item As expected given the frequency dependence of the noise visible on Fig.~\ref{fig_noise}, the standard deviation is smaller in spin V3. Furthermore it  logically depends on the duration of the analysis segment: it is shown in \cite{chhuncqg5}, that when normalised to the same duration (or equivalently, expressed in terms of PSD) all sessions with the same SU and the same spin have quite similar noise. 
\item As expected before the launch, {SUREF provided more accurate data than SUEP}, mainly thanks to its heavier outer test-mass.  Indeed the ratio area over mass is a driving factor of the error budget.
\end{enumerate}

In terms of {the} E\"otv\"os parameter, these results are statistically compatible with a null value: all absolute values are smaller than $2 \sigma$ except for sessions 294 and 380 where they exceed $3 \sigma$ but remain below $5 \sigma$. Accounting for systematic errors (see Table \ref{tab-systematics-SUREF}) further downplays the significance of these values. The value of $\sigma$ depends on the segment and in particular of its length but is typically $10^{-14}$ or smaller for the longest segments of 120 orbits.

\subsection{{Robustness tests}}
\label{sec:impact-param}
We conducted a series of complementary analyses to test the robustness of our results and quantify how they are impacted by the settings of the analysis and by some terms considered as negligible according to the specifications. For reason of CPU time, all these tests have been performed using the ADAM software.

\subsubsection{Impact of the frequency bandwidth.}
\label{sec:impact-freq-bandw}

As explained above and in \cite{bergecqg7}, we select a frequency band around $f_{\rm EP}$ and $2 f_{\rm EP}$ to compute the parameters of the model using a weighted least-squares estimation in the frequency domain; we have checked that dividing the used bandwidth around these frequencies by a factor 4 does not change the values of the parameters by more than a few percent (a few $10^{-16}$ for $\delta_x $); however the associated standard deviation can change by up to 10 \% which is understandable because if we reduce the number of data (in the frequency domain) its estimation is less precise. In the same spirit, if we no longer use a selected band of frequency but the whole domain, the parameters are not noticeably modified and the estimated standard deviation generally increases from 10 to 20\%; indeed, in this case, we could have unmodeled high frequency effects which contribute to increasing the standard deviation. This is also probably why $\sigma$ estimated by M-ECM  (which implicitly uses the whole frequency domain) is generally slightly larger than the one obtained with ADAM.

\subsubsection{Impact of the degree of the fitted polynomial.}
\label{sec:impact-degree-fitted}
In the {actual} analysis, we estimate a polynomial of degree 3  in order to absorb  the low frequency trend due to temperature variations \cite{bergecqg7}.
We have also tested other degrees, from 1 to 5.  While there is no significant modification with degrees 2, 4 or 5 the standard deviation is slightly increased when using a degree 1.

We have also compared different strategies to estimate the coefficients of the polynomial:
\begin{enumerate}
\item a prior estimation (without weighting) in the time domain,
\item a prior estimation (with weighting according to the estimated power spectral density of the noise) in the Fourier domain (using the whole spectrum),
\item estimation at the same time as the other parameters using the whole spectrum but the frequency bands around $f_{\rm EP}$ and $2 f_{\rm EP}$.
\end{enumerate}
The final results for the estimated E\"otv\"os parameter and the associated standard deviation are fully equivalent: the differences are much smaller than the error and than our ideal objective of $10^{-15}$.

\subsubsection{Impact of the angular velocity and angular acceleration.}
\label{sec:impact-angul-veloc}
Angular velocity (included in the matrix $\left[{\mathbf S}\right]$) and angular acceleration are used in the {actual} analysis to correct the measured linear acceleration from gradient effects { due to offcentring} ${\vv \Delta}$. The stability of the attitude of MICROSCOPE has been specified to limit the effect of this gradient at the $f_{\rm EP}$ and $2f_{\rm EP}$ frequencies. To check that this is indeed the case, we have analysed the data without these corrections. {Again, this resulted in negligible changes in the estimated values of the E\"otv\"os parameter, less than 5\% of its standard deviation except for 3 segments for which this was about 10\%.}

\subsubsection{Impact of the residual variations of position of the test masses.}
\label{sec:impact-resid-vari}
Fundamentally, the full dynamical equation of the test masses includes terms taking into account the motion of the masses with respect to the satellite (see Eq.~(1) in  \cite{rodriguescqg1}): the kinematic  acceleration $\ddot{\vv{\Delta}} $ and the Coriolis acceleration $2\vv{\Omega} \times{}\dot{\vv{\Delta}} $. Since the principle of the accelerometer is to nullify the displacement of the test masses, these terms are not included in the {actual} analysis. Nevertheless position measurements are available in the telemetry data. Due to the limited quantity of data that can be transmitted on ground, they are sampled at 1 Hz. Thus, in order to compute the previous terms and correct the electrostatic acceleration {we have both interpolated the position at 4 Hz and derived them numerically.} The modification of the estimated $\delta_x$ amounts to a few percents of its standard deviation.

\subsection{Testing the whole process with a simulated EP signal}
\label{sec:test-whole-proc}

  It is essential to check that  the global analysis  process, including  the detection-elimination of the glitches, is able to preserve and retrieve a potential EP-violation signal. To this aim we added a fake signal to the real measurements before any preprocessing and applied our complete chain of analysis going from the masking described in Sect. \ref{sec:detect-elim-glitch} to the estimation of the E\"otv\"os parameter as explained above. More precisely, we conducted two series of tests with two levels of added simulated signal: one corresponding to an E\"otv\"os parameter of  $3.4\times{}10^{-14}$ (which is large compared to the objective of MICROSCOPE) and a second corresponding to an E\"otv\"os parameter of  $3.4\times{}10^{-15}$ (which is more or less the limit of detection for SUEP as confirmed by Eq.~\ref{eq:13}). We then performed the analysis of these data and subtracted the E\"otv\"os parameter estimated with the original data. Tables \ref{tab-fakeEP_SUREF} and \ref{tab-fakeEP_SUEP} show these differences which can be compared to the simulated signal.
The standard deviations corresponding to these {analyses} are very close to those of the initial analysis (Tables \ref{tab-estimations-SUREF} and \ref{tab-estimations-SUEP}) and are not repeated here.
Tables \ref{tab-fakeEP_SUREF} and \ref{tab-fakeEP_SUEP} show also the bias (i.e. the difference quoted above minus the simulated value) divided by the standard deviation of the E\"otv\"os parameter. The absolute value of this ratio is smaller than 2\% for all segments of the SUEP and for most of the segments of the SUREF. The  worst case is for segment 778-1 (SUREF) with a relative error of 18\% for a fake signal of $3.4\times{}10^{-14}$; but even in this case the absolute error is smaller than $10^{-15}$.

\begin{table}
\caption{\label{tab-fakeEP_SUREF}Estimation of a simulated fake EP signal to test the robustness of the analysis process {for} SUREF.} 
\begin{indented}
\lineup
\item[]\begin{tabular}{@{}lcccc}
\br                              
 $\0\0$                & \centre{2}{Fake E\"otv\"os parameter  $3.40\times{}10^{-15} $ }&\centre{2}{Fake E\"otv\"os parameter $34.00\times{}10^{-15} $} \cr
\ns
$\0\0$Segment       &\crule{2}                                                                   & \crule{2}                                                                      \cr
$\0\0$number   &    Estimated  ($\times{}10^{15}$)  &                      bias / $\sigma$   &    Estimated  ($\times{}10^{15}$)  &   bias / $\sigma$     \cr 
\mr
\0\0{120-1}        &{3.41}    & \phantom{-}0.00         &{34.01}     & \phantom{-}0.00     \cr
\0\0{120-2}        &{3.45}    & \phantom{-}0.01         &{34.00}     & \phantom{-}0.00      \cr
\0\0{174}            &{3.40}    & \phantom{-}0.00          &{33.99}     & \phantom{-}0.00      \cr
\0\0{176}            &{3.44}    & \phantom{-}0.01          &{34.04}      & \phantom{-}0.01     \cr
\0\0{294$^*$}     & {3.34}   & {-0.03}                         & {33.97}    & -0.01                        \cr
\0\0{376-1}        &{3.46}    & \phantom{-}0.01          &{33.93}    & -0.01                         \cr
\0\0{376-2}        &{3.31}    & -0.02                            &{33.96}    & -0.01                        \cr
\0\0{380-1$^*$} & {3.41}    & \phantom{-}0.00         &{34.07}    & \phantom{-}0.03      \cr
\0\0{380-2$^*$} &{3.23}    & -0.06                            &{33.81}     & -0.07                       \cr
\0\0{452}            &{3.43}    & \phantom{-}0.01          &{34.04}    & \phantom{-}0.01       \cr
\0\0{454}            &{3.37}    & -0.01                            &{33.98}     & -0.01                       \cr
\0\0{778-1}        &{3.39}    & \phantom{-}0.00          &{33.18}     & -0.18                       \cr
\0\0{778-2}        &{3.44}   & \phantom{-}0.01           &{34.09}      & \phantom{-}0.02     \cr
\br
\end{tabular}
\end{indented}
\end{table}

\begin{table}
\caption{\label{tab-fakeEP_SUEP}Estimation of a simulated fake EP signal to test the robustness of the analysis process {for} SUEP.} 
\begin{indented}
\lineup
\item[]\begin{tabular}{@{}lcccc}
\br                              
 $\0\0$                & \centre{2}{Fake E\"otv\"os parameter  $3.40\times{}10^{-15} $ }&\centre{2}{Fake E\"otv\"os parameter $34.00\times{}10^{-15} $} \cr
\ns
$\0\0$Segment       &\crule{2}                                                                   & \crule{2}                                                                      \cr
$\0\0$number   &    Estimated  ($\times{}10^{15}$)      &                      Error / $\sigma$   &    Estimated    ($\times{}10^{15}$) &   Error / $\sigma$     \cr 
\mr
\0\0{210$^*$}    & {3.22}  &-{0.01}                    & {34.14}  & \phantom{-}{0.01}   \cr
\0\0{212$^*$}    &{3.36}   &\phantom{-}{0.00}  & {33.95}  & \phantom{-}{0.00}   \cr
\0\0{218$^*$}    & {3.51}  & \phantom{-}{0.01} & {34.13}  &  \phantom{-}{0.02}     \cr
\0\0{234$^*$}    & {3.39}  & \phantom{-}{0.00}& {33.85}  & -{0.02}                      \cr
\0\0{236$^*$}    & {3.41}  &\phantom{-}{0.00}& {33.99}  & \phantom{-}{0.00}    \cr
\0\0{238$^*$}    & {3.42}  &\phantom{-}{0.00} &{34.00}   & \phantom{-}{0.00}   \cr
\0\0{252$^*$}    & {3.37}  &\phantom{-}{0.00} &{33.97}   & \phantom{-}{0.00}    \cr
\0\0{254$^*$}    & {3.37}  &\phantom{-}{0.00} & {33.98}  & \phantom{-}{0.00}    \cr
\0\0{256$^*$}    &{3.51}   & \phantom{-}{0.01} & {33.88}  & -{0.02}                     \cr
\0\0{326-1$^*$}& {3.42}  &\phantom{-}{0.00} &{34.23}   & \phantom{-}{0.02}     \cr
\0\0{326-2$^*$}&{3.33}   &\phantom{-}{0.00} & {33.77}  & -{0.02}                      \cr
\0\0{358$^*$}    &{3.41}   &\phantom{-}{0.00}  &{34.06}   & \phantom{-}{0.01}   \cr
\0\0{402}          &{3.54}   &\phantom{-}{0.00}  & {33.93} & \phantom{-}{0.00}   \cr
\0\0{404$^*$}    & {3.38}  &\phantom{-}{0.00} & {34.01}  & \phantom{-}{0.00}   \cr
\0\0{406$^*$}    & {3.37}  &\phantom{-}{0.00} &{34.00}   & \phantom{-}{0.00}  \cr
\0\0{438}          &{3.45}   &\phantom{-}{0.00}  & {34.19} & \phantom{-}{0.01} \cr
\0\0{442}          &{3.45}   &\phantom{-}{0.00} & {34.02}  & \phantom{-}{0.00} \cr
\0\0{748}          &{3.40}   &\phantom{-}{0.00} & {34.08}  & \phantom{-}{0.00}  \cr
\0\0{750$^*$}    & {3.95}  &\phantom{-}{0.01} & {34.66} &\phantom{-}{0.02}  \cr
\br
\end{tabular}
\end{indented}
\end{table}

\subsection{Systematic errors}
\label{sec:systematic-errors}

The evaluation of systematic errors is a major topic addressed during the preparation and specification of the mission and also since the end of the mission in 2018 using the actual measurements and characterisation of the experiment.
Systematic errors are estimated in detail in \cite{hardycqg6} (see Table 15 therein for a summary). They can be divided into seven main contributors, the maximum impact of each was estimated at the $f_{\rm EP}$ frequency.
\begin{enumerate}
\item Residual gravitational effects either due to imperfect knowledge of the Earth's gravity field and of the position and orientation of the instrument, or due to local effects coming from the satellite or the instrument itself (first three rows of Table 15 in \cite{hardycqg6}). Those residuals are due either to errors in the correction of the Earth's gravity gradients after calibration of the offcentrings, or to errors in the estimation of local gravity fluctuations (performed with finite element analysis before the launch).
\item {Clock errors: the most stringent requirement in terms of time stamping comes from the need to compute the gravity gradient tensor with the correct position and orientation of the satellite, in order to correct the effects of the gravity gradient. Since the dominant contributions of the gravity gradient are at DC and  $2 f_{\rm EP}$ this contrains the absolute clock errors at $f_{\rm EP}$  and $3 f_{\rm EP}$\footnote{The combinations  $f_{\rm EP}$ with DC,  $f_{\rm EP}$ with $2f_{\rm EP}$ and $3f_{\rm EP}$ with $2f_{\rm EP}$ can all produce effects at  $f_{\rm EP}$.}; they have been specified  to be smaller than 1\,ms,  leading to an error smaller than $2\times{}10^{-16}$ ms$^{-2}$ on the gravity gradient effect. The maximum effective errors were 2.1\,$\mu$s and 0.7\,$\mu$s respectively at $f_{\rm EP}$ and $3f_{\rm EP}$ in inertial pointing and even smaller in spin mode. The maximum total error (including bias, harmonic errors and drift) with respect to UTC was specified to be smaller than 50\,ms and was always smaller than 41.3\,ms. In order to limit the effect of the drift, the on-board clock was regularly synchronized (outside the scientific sessions).}
\item Uncorrected inertial effects due to imperfect estimation of angular velocity and angular acceleration. This contribution is computed via the estimated performance of the DFACS \cite{robertcqg3, rodriguescqg4}, which allows estimation of the angular error contribution at $f_{\rm EP}$ for each session.
\item Time variation of the instrument parameters (satellite pointing, common mode test-mass alignment and angular-to-linear acceleration couplings). These variations are not due to temperature variation  at $f_{\rm EP}$ (which are taken into account in the thermal effect below). They are computed by combining the  satellite alignment variation issued from the DFACS control with the residual continuous differential acceleration at the differential measurement output.
\item The drag-free control residuals.  In complementarity with the previous item, instrument parameters (estimated from calibration sessions) are assumed constant during a given session. The DFACS-related systematic error is computed from the residual accelerations at $f_{\rm EP}$ issued from the DFACS performance report.
\item Magnetic effects.  They are computed with a finite element model partially adjusted to measurements made on the instrument magnetic shielding and based on the knowledge of all electronics units characteristics.
\item thermal effects induced by the tiny variations of temperature at the $f_{\rm EP}$ frequency. These are the major contributors to the systematic error budget. Their estimation results from a detailed analysis of the instrument and of the satellite thermal behaviours. It should be noted that the temperature variations at $f_{\rm EP}$ in the SU were reduced to a fraction of $\mu {\rm K}$ in the worst case and helped to strongly reduce the thermal contributor compared to that of \cite{touboul17}.
\item Non-linearities, described by a quadratic term in the measurement equation. Note that \Eref{eq_xacc} ignores this term because it is not used in the data process but only in a posteriori analysis of errors \cite{hardycqg6}.
\end{enumerate}

According to \cite{hardycqg6}, the last two terms are dozens of times larger than the others for SUEP. In the case of SUREF, non-linear effects are smaller and thermal effects dominate all the others. These various effects are unlikely to be correlated and are quadratically added at the end.

The main systematics{ (in terms of E\"otv\"os parameter)} as well as their total contribution {${\cal S}_l$ for each segment $l$} are {summarised} in Tables \ref{tab-systematics-SUREF} and \ref{tab-systematics-SUEP}.
Following~\cite{hardycqg6}, {${\cal S}_l$ is computed according to ${\cal S}_l=\frac{1}{g}\sqrt{\sum_k\left(\Gamma_{k,l}^{(d)} \right)^2}$ where $\Gamma_{k,l}^{(d)} $ are the maximum systematic errors of acceleration corresponding to each source $k$, and $g\simeq 7.9\times{}10^{-15}$ ms$^{-2}$ is the gravity acceleration for MICROSCOPE.}
The standard deviation issued from the least-squares regression with ADAM presented in Section \ref{sec:results} is also recalled for comparison. This shows that stochastic errors are clearly dominant for SUEP and marginally dominant for SUREF.

\begin{table}
\caption{\label{tab-systematics-SUREF}Main systematic effects for SUREF sessions: thermal effects and non-linear (quadratic) effects which are dominant and total computed as the quadratic sum of all effects detailed in  \cite{hardycqg6}. The last column repeats, for comparison, the stochastic error obtained with ADAM presented in Table \ref{tab-estimations-SUREF}. All values are given in equivalent of $10^{-15}$ for the E\"otv\"os parameter (for example systematic errors in ms$^{-2}$ have been divided by $10^{-15}\,g\simeq 7.9\times{}10^{-15}$ ms$^{-2}$). Segments marked by an asterisk $^*$ correspond to sessions with spin V3 and the others to sessions with spin V2.} 

\begin{indented}
\lineup
\item[]\begin{tabular}{@{}lrrrr}
\br                              
$\0\0$Segment & Thermal &Quadratic& Total           & Stochastic \cr 
$\0\0$number   &effect     &effect       &systematics&error            \cr 
\mr
\0\0{120-1}        &{5.4}           &{0.1}          &{5.4}          &{15.9}       \cr
\0\0{120-2}        &{5.4}           &{0.1}         &{5.4}          &{8.3}        \cr
\0\0{174}            &{4.0}           &{0.4}         &{4.0}          &{4.2}        \cr
\0\0{176}            &{4.7}           &{0.4}         &{4.7}          &{4.5}        \cr
\0\0{294$^*$}     & {2.0}          & {0.4}        &  {2.0}        & {2.1}       \cr
\0\0{376-1}        &{3.7}           &{0.4}         &{3.7}          &{6.5}        \cr
\0\0{376-2}        &{3.7}           &{0.4}         &{3.7}          &{5.8}        \cr
\0\0{380-1$^*$} & {0.7}          & {0.4}        &{0.8}          &{2.4}       \cr
\0\0{380-2$^*$} &{0.7}           &{0.4}         &{0.8}          &{2.8}       \cr
\0\0{452}            &{2.1}           &{0.5}         &{2.1}          &{4.1}        \cr
\0\0{454}            &{2.7}           &{0.4}         &{2.7}          &{2.8}        \cr
\0\0{778-1}        &{3.0}           &{0.3}         &{3.0}          &{4.7}        \cr
\0\0{778-2}        &{3.0}           &{0.3}         &{3.0}          &{5.5}        \cr
\br
\end{tabular}
\end{indented}
\end{table}

\begin{table}
  \caption{\label{tab-systematics-SUEP}Same as Table \ref{tab-systematics-SUREF} but for SUEP.}

\begin{indented}
\lineup
\item[]\begin{tabular}{@{}lrrrr}
\br                              
$\0\0$Segment & Thermal &Quadratic& Total           & Stochastic \cr 
$\0\0$number   &effect     &effect       &systematics&error            \cr 
\mr
\0\0{210$^*$}    & {1.7}  &{0.8}  & {1.8}  & {13.1}   \cr
\0\0{212$^*$}    &{0.6}   &{0.9}  & {1.0}  & {11.9}   \cr
\0\0{218$^*$}    & {0.8}  & {0.7} & {1.1}  & {8.1}     \cr
\0\0{234$^*$}    & {0.8}  & {0.7} & {1.0}  & {8.3}     \cr
\0\0{236$^*$}    & {1.0}  & {0.7} & {1.2}  & {6.6}     \cr
\0\0{238$^*$}    & {1.0}  & {0.7} &{1.2}   & {6.4}    \cr
\0\0{252$^*$}    & {0.8}  &{0.7}  &{1.1}   & {7.3}    \cr
\0\0{254$^*$}    & {1.3}  & {0.8} & {1.5}  & {7.0}    \cr
\0\0{256$^*$}    &{0.8}   & {0.7} & {1.1}  & {7.4}    \cr
\0\0{326-1$^*$}& {0.8}  &{1.3}  &{1.6}   & {9.6}     \cr
\0\0{326-2$^*$}&{0.8}   & {1.3} & {1.6}  & {13.5}  \cr
\0\0{358$^*$}    &{0.8}   &{0.7}  &{1.1}   & {10.9}   \cr
\0\0{402}          &{7.3}  & {0.7} & {7.3}& {43.6} \cr
\0\0{404$^*$}    & {0.7}  & {0.7} & {1.0}  & {6.7}    \cr
\0\0{406$^*$}    & {3.1}  & {0.8} &{3.2}   & {14.9}  \cr
\0\0{438}          &{5.4}  & {0.6} & {5.5}& {24.6}  \cr
\0\0{442}          &{7.2}  & {0.6} & {7.3}& {19.1}  \cr
\0\0{748}          &{7.2}  & {0.6} & {7.3}& {24.6}  \cr
\0\0{750$^*$}    & {7.2}  & {0.7} & {7.3} & {38.4}  \cr
\br
\end{tabular}
\end{indented}
\end{table}

\section{Estimation of the E\"otv\"os parameter using the combination of sessions}
\label{sec:estim-eotv-param-1}
The results from the analysis of the {individual}  segments show that stochastic errors are larger than systematic errors. In order to  improve the signal to noise ratio we have gathered, in a global analysis, all segments included in Tables \ref{tab-estimations-SUREF} and \ref{tab-estimations-SUEP}. The polynomial coefficients $\alpha_j$ and the offcentring $\Delta'_{x} $ and $\Delta'_{z}$ are specific to each segment, but the parameters $\delta_x $ and $\delta_z$ are common. This has been achieved in the Fourier domain (with the ADAM software) as described in \cite{bergecqg7} {and recalled in} \ref{sec:lest-squaress-fourie}. {Due to excessively large gaps between segments, we can not simply cumulate the corresponding measurements without being overwhelmed by leakage effects, even when applying dedicated algorithms like M-ECM. Instead, we apply a Discrete Fourier Transform to each segment and cumulate the transformed equations in the Fourier domain as detailed in} \ref{sec:lest-squaress-fourie}.

As a result we get for SUREF
\begin{equation}
  \label{eq:3}
  \delta_x =( 0.0\pm{}1.1)\times{}10^{-15},
\end{equation}
and for SUEP
\begin{equation}
  \label{eq:5}
  \delta_x=( -1.5\pm{}2.3)\times{}10^{-15},
\end{equation}
where the errors given above are statistical errors at 1 $\sigma$. As was to be expected from the results of individual segments, SUEP suffers from a larger statistical error than SUREF despite the larger number of segments used in the combined solution.

{
  In the same spirit, we have analysed the cumulated segments containing fake signals as described in Sect. \ref{sec:test-whole-proc}. For the SUREF the estimated increment on the E\"otv\"os ratio  is $3.38\times{}10^{-15}$ when $3.40\times{}10^{-15}$ was simulated and  $34.01\times{}10^{-15}$ when $34.00\times{}10^{-15}$ was simulated. For the SUEP the results {are} respectively $3.37\times{}10^{-15}$ and  $33.99\times{}10^{-15}$.
 }

\begin{figure} 
\center
\includegraphics[width=0.48\textwidth]{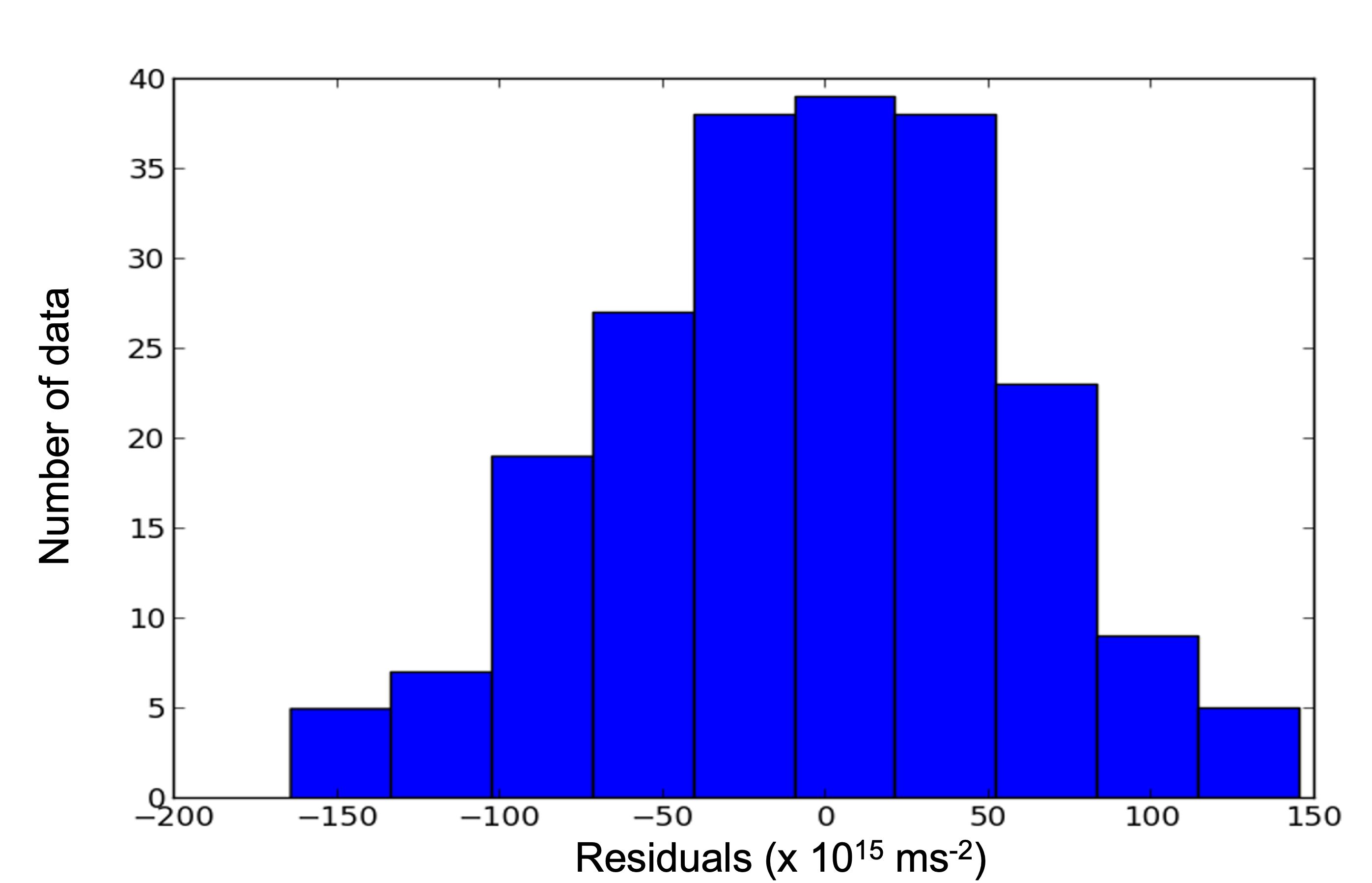}
\includegraphics[width=0.49\textwidth]{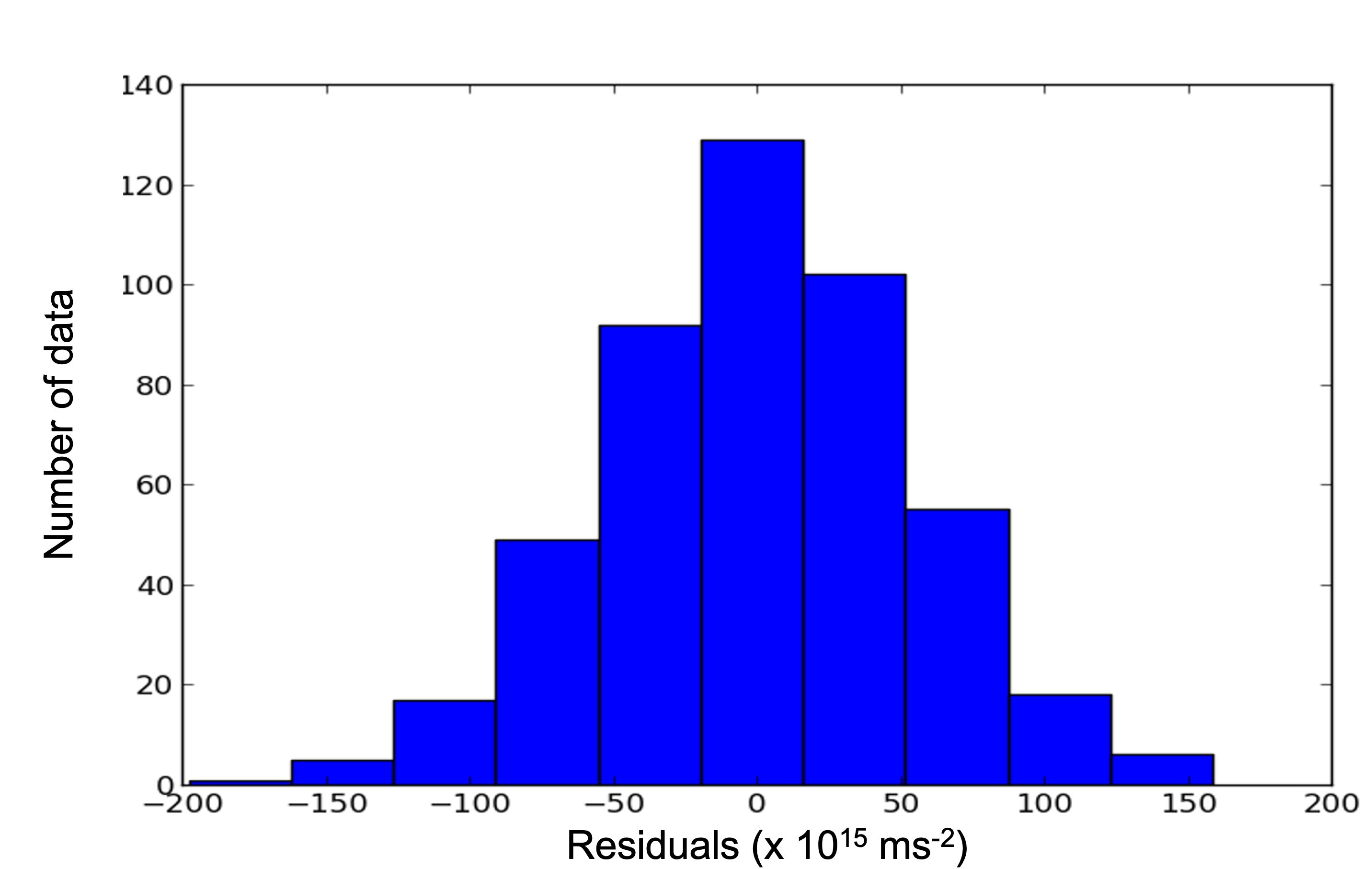}
\caption{Histograms of the residuals, in the frequency band around $f_{\rm EP}$, of the measured acceleration after fitting of the model (\ref{eq:2}) in the Fourier domain for SUREF (left panel) and SUEP (right).}
\label{fig_histo}       
\end{figure}

Fig.~\ref{fig_histo} shows the  histograms of the weighted residual accelerations in the frequency band around $f_{\rm EP}$ after estimation of the parameters and subtraction  of the model \eqref{eq:2}. {We have checked that they are compatible with a Gaussian statistics.}
Note that if instead of gathering all measurements to provide the global solution, we compute the weighted mean of the solutions for individual segments $l$ as
\begin{equation}
  \label{eq:8}
    \delta_{x,{\rm M}}=\frac{\displaystyle\sum_l\frac{\delta_{x,l}}{\sigma_l^2}}{\displaystyle\sum_l\frac{1}{\sigma_l^2}},
\end{equation}
{and the associated variance} 
\begin{equation}
  \label{eq:15}
  \sigma^2_{{\rm M}}=\frac{1}{\displaystyle\sum_l\frac{1}{\sigma_l^2}},
\end{equation}
we get the very similar results $ \delta_{x,{\rm M}} =( -0.4\pm{}1.1)\times{}10^{-15}$ for SUREF and $ \delta_{x,{\rm M}} =( -1.8\pm{}2.2)\times{}10^{-15}$ for SUEP. This {is expected if the observations of the different segments are sufficiently independent. The same weighting is used to combine systematic errors ${\cal S}_l$ associated to individual segments (given in column 4 of Tables~}\ref{tab-systematics-SUREF} and \ref{tab-systematics-SUEP}) {in order to get  the systematic error $ {\cal S}_{{\rm M}}$ associated to the global solution}\cite{hardycqg6}:
\begin{equation}
  \label{eq:9}
  {\cal S}_{{\rm M}}=\frac{\displaystyle\sum_l\frac{  {\cal S}_l}{\sigma_l^2}}{\displaystyle\sum_l\frac{1}{\sigma_l^2}},
\end{equation}
which leads to the systematics $ {\cal S}_{{\rm M}}=2.3\times{}10^{-15}$ for SUREF and $ {\cal S}_{{\rm M}}=1.5\times{}10^{-15}$ for SUEP.

{Putting all together, and remembering that  the conventional E\"otv\"os parameter $\eta$ {can be practically identified} to the parameter  $\delta_{x}$ measured in this experiment,  we end up for SUREF with}
\begin{equation}
  \label{eq:10}
  \eta({\rm{Pt, Pt}}) \simeq\delta({\rm{Pt, Pt}})=[0.0\pm{}1.1{\rm (stat)}\pm{}2.3{\rm (syst)}] \times{}10^{-15}\ \rm{at} ~1\sigma.
\end{equation}
As a null E\"otv\"os parameter is expected in this case, this result gives a good indication  that there is no important anomaly in the whole chain going from the measurements to the analysis and including the modelling.

We finally get for SUEP
\begin{equation}
  \label{eq:13}
  \eta({\rm{Ti, Pt}}) \simeq\delta({\rm{Ti, Pt}})=[-1.5\pm{}2.3{\rm (stat)}\pm{}1.5{\rm (syst)}] \times{}10^{-15} \ \rm{at} ~1\sigma.
\end{equation}

{This final result  indicates that there is no visible violation of the WEP in the  full data of the MICROSCOPE mission.}

\section{Conclusion} \label{sect_conclusion}

We have analysed the measurements provided by the payload T-SAGE {flown} on the MICROSCOPE satellite: these are the differences of accelerations of two test-masses made of the same material (PtRh10) for the sensor SUREF and two  test-masses  made of different material (PtRh10 for the inner mass, Ti alloys for the outer mass) for the sensor SUEP.
This involves 13 segments (i.e. sequences of continuous measurements sampled at 4 Hz)  totalling 598 orbits for  SUREF  and 19 segments totalling 1362 orbits for SUEP. This represents accumulated free falls in the Earth's gravity field of about 41 days for SUREF and 94 days for SUEP.
{In the data analysis we have compared the measurements to a model including many effects, in particular those of the gravity gradients and of the gradient of inertia due to the tiny difference of positions of the test masses, together with a hypothetical signal of violation of the Equivalence Principle. Before these computations we have corrected and calibrated the instrumental parameters estimated during the dedicated calibration sessions. We have moreover detected and discarded the measurements  affected by glitches. As this process breaks the regularity of the sampling, we had to use appropriate algorithms in order to prevent the effects of leakage.}

In a first step the analysis has been performed separately on single segments using two different methods: M-ECM operating in the time domain which has been designed for optimal estimation using irregularly sampled data and is also able to reconstruct the most likely data where they are missing, and ADAM operating in the frequency domain. The two methods give consistent {estimates of the parameter $\delta=\eta+{\cal O}\left(\eta^2 \right)$}.
{In particular, the values estimated with the SUEP accelerometer on the different segments are consistent with 0 at less than 2 $\sigma$ for most of them, 2.2 $\sigma$ for one of them. This distribution is compatible with Gaussian statistics.}
 The value of $\sigma$ depends on the segment and in particular on its length but is typically $10^{-14}$ or smaller for the longest segments of 120 orbits. The systematic errors, analysed in \cite{hardycqg6} for the same segments, are significantly smaller.

 {In a final step we have gathered the data coming from all the segments for each SU in a single analysis, with  the aim of obtaining  the best signal to noise ratio from our full set of data. 
 This led again to no detection of  violation of the WEP since we obtained   $\eta({\rm{Ti, Pt}}) =[-1.5\pm{}2.3{\rm (stat)}\pm{}1.5{\rm (syst)}] \times{}10^{-15}$ for the SUEP.   The result obtained for the SUREF,  $\eta({\rm{Pt, Pt}}) =[0.0\pm{}1.1{\rm (stat)}\pm{}2.3{\rm (syst)}] \times{}10^{-15}$ confirmed the absence of bias in the whole analysis,  a null value being expected in this case.}   It is common \cite{jcgm, wagner12}  to add quadratically the statistic{al} and systematic errors. Doing this we conclude that the MICROSCOPE experiment does not see evidence for any difference of free fall between titanium and platinum test masses {at a level of sensitivity of} $2.7\times{}10^{-15}$.  This represents an improvement of almost two orders of magnitude with respect to the constraint before the launch of MICROSCOPE. 

{Although this new upper bound on the WEP allows for improved bounds on beyond-GR models (see e.g.}~\cite{berge18, fayet18, fayet19} {for bounds obtained after the first MICROSCOPE results}~\cite{touboul19, touboul17}), {the challenges faced by fundamental physics remain as pressing as ever and call for still more precise experiments. New tests in space could improve MICROSCOPE's state-of-the art measurement by two orders of magnitude in the next decades}~\cite{batt21}.

\ack

The authors express their gratitude to all the different services involved in the mission partners and in particular CNES, the French space agency in charge of the satellite. This work is based on observations made with the T-SAGE instrument, installed on the CNES-ESA-ONERA-CNRS-OCA-DLR-ZARM MICROSCOPE mission. ONERA authors’ work is financially supported by CNES and ONERA fundings.
Authors from OCA, Observatoire de la C\^ote d’Azur, have been supported by OCA, CNRS, the French National Center for Scientific Research, and CNES. ZARM authors’ work is supported by the DLR, German Space Agency,  with funds of the BMWi (FKZ 50 OY 1305 and FKZ 50 LZ 1802) and by the Deutsche Forschungsgemeinschaft DFG (LA 905/12-1). The authors would like to thank the Physikalisch-Technische Bundesanstalt institute in Braunschweig, Germany, for their contribution to the development of the test-masses with funds of CNES and DLR.

\appendix

\section{{Least-squares regression in the Fourier domain}}
\label{sec:lest-squaress-fourie}

We recall here the main steps of the procedure described in  \cite{bergecqg7}.

The set of measurement equations \eqref{eq:2} applied at each time of measurement  $t_i$ can be written as a linear system
\begin{equation}
  \label{eq:1}
  {\mathbf Y} = {\left[{\mathbf A}\right]} {\boldsymbol{\theta}}+{\mathbf n},
\end{equation}
where ${\mathbf Y} $ is the vector of $N$ measurements, ${\boldsymbol{\theta}}$ is a vector of $q$ unknown parameters to estimate (i.e. $\alpha_j$, $\delta_x$, $\delta_z$, $\Delta'_{x}$ and $\Delta'_{z}$),  $\left[{\mathbf A}\right]$ is the design matrix and ${\mathbf n}$ is the noise vector.
Since the model is  linear with respect to the estimated parameters, the columns of  $\left[{\mathbf A}\right]$ simply correspond to the signal associated to each parameter, sampled at the epochs of the measured acceleration.

The $N$ measurements are assumed to be regularly sampled at a frequency $f_{\rm e}$ over a duration $T$. In order to solve the problem in the Fourier domain, we take the Fourier transform of Eq.~(\ref{eq:2}). To this aim, we make use of the DFT operator $\left[{\mathbf F}\right]$. The DFT operator being unitary, the signal energy content is preserved by the transformation.
The new system can be simply written
\begin{equation}
  \label{eq:7}
    \hat{\mathbf Y} = {\left[\hat{\mathbf {A}}\right]} {\boldsymbol{\theta}}+\hat{\mathbf n}.
 \end{equation}
The original quantities being real, the new system can be reduced to $N$ useful real equations.
These new equations can be grouped by pair (related to real and
imaginary parts of the DFT), corresponding to frequencies
$f_k=\frac{k}{T}, k=1\cdots\lfloor\frac{N-1}{2}\rfloor$.

 Since each measurement projected in the Fourier domain can be
associated to a discrete frequency, the corresponding weight is
\begin{equation}
  \label{eq:4}
  w(f_k)=\frac{1}{\sqrt{\gamma(f_k)}},
\end{equation}
where $\gamma(f_k)$ is the PSD of the noise at the
frequency $f_k$.

As shown in Ref.~\cite{rodriguescqg1}, the MICROSCOPE mission was designed to concentrate useful signal at specific frequencies (i.e., gravity acceleration peaks at $f_{\rm EP}$, the gravity gradient signal at $2f_{\rm EP}$ and calibration signals at $f_{\rm cal}$). This is so true in the real data that a very simple analysis
such as synchronous detection could lead to reasonable results.
However, we use a more flexible method: we limit our least-squares inversion to the bands of frequency containing the relevant signals. In practice, this is equivalent to extracting a subsystem of Eq.~\eqref{eq:7} by selecting the  relevant equations to get the truncated system
\begin{equation}
  \label{eq:12}
    {\left[\hat{\mathbf {A}}_r\right]} {\boldsymbol{\theta}}+\hat{\mathbf n} =\hat{\mathbf Y}_r.
  \end{equation}

 One can cumulate the data from disjoint segments by just gathering the corresponding matrices and vectors:
\begin{equation}
  \label{eq:6}
  \left[\hat{\mathbf A}\right]=
  \begin{bmatrix}
     {\left[\hat{\mathbf A}_1\right]} \\
 {\left[\hat{\mathbf A}_2\right]} \\
\vdots \\
 {\left[\hat{\mathbf A}_m\right]} 
  \end{bmatrix},
\quad
\hat{\mathbf n}=
  \begin{bmatrix}
     \hat{\mathbf n}_1 \\
 \hat{\mathbf n}_2\\
\vdots \\
 \hat{\mathbf n}_m
  \end{bmatrix},
\quad
\hat{\mathbf Y}=
  \begin{bmatrix}
     \hat{\mathbf Y}_1 \\
 \hat{\mathbf Y}_2\\
\vdots \\
 \hat{\mathbf Y}_m
  \end{bmatrix}
\end{equation}
where $m$ is the number of sessions considered. Then an appropriately weighted least-squares technique can be used to solve for the concatenated system, the weight associated to each frequency for each segment being chosen according to \eqref{eq:4}.

In the simplest cases, all parameters are common to all segments.
If some parameters are specific to each segment (as could be the case for the polynomial coefficients or for the offcentring), the corresponding column of the design matrix related to the other segments is simply set to zero.

\section {List of acronyms and abbreviations}
\label{sec:acron}
{ADAM: Accelerometric Data Analysis for MICROSCOPE} \\
DC: Direct Continuous \\
DFACS: Drag-Free and Attitude Control System \\
DFT: Discrete Fourier Transform\\
EP: Equivalence Principle\\
IS: Inertial Sensor\\
M-ECM: Modified Expectation Conditional Maximisation\\
MLI: Multi-Layer Insulation\\
PSD: Power Spectrum Density\\
SU: Sensor Unit\\
SUEP: Sensor Unit for the Equivalence Principle test\\
SUREF: Sensor Unit for Reference\\
T-SAGE: Twin Space Accelerometer for Gravity Experiment is the name of the payload \\
{UTC: Universal Time Coordinated} \\
WEP: Weak  Equivalence Principle

\section*{References}
\bibliographystyle{iopart-num}
\bibliography{biblimscope}

\end{document}